# Accurate Parametric Inference for Small Samples

Alessandra R. Brazzale and Anthony C. Davison

*Abstract.* We outline how modern likelihood theory, which provides essentially exact inferences in a variety of parametric statistical problems, may routinely be applied in practice. Although the likelihood procedures are based on analytical asymptotic approximations, the focus of this paper is not on theory but on implementation and applications. Numerical illustrations are given for logistic regression, nonlinear models, and linear non-normal models, and we describe a sampling approach for the third of these classes. In the case of logistic regression, we argue that approximations are often more appropriate than 'exact' procedures, even when these exist.

## 1. INTRODUCTION

Monte Carlo inference has developed remarkably over the last 30 years. Bootstrap procedures (Efron (1979)) are used for a wide range of problems (Efron and Tibshirani (1993), Davison and Hinkley (1997)). Markov chain Monte Carlo simulation has transformed Bayesian modelling (Robert and Casella (2004)). The combination of iterative simulation with importance sampling and improved algorithms for full enumeration of discrete sample spaces has had a strong impact on the analysis of contingency tables (Forster, McDonald and Smith (1996), Smith, Forster and McDonald (1996), Diaconis and Sturmfels (1998), Mehta, Patel and Senchaudhuri (2000)). More recently there has been a rise in Bayesian nonparametric modelling (Denison et al. (2002)), which parallels the use of the bootstrap for nonparametric frequentist inference. All these techniques use simulation to avoid tailoring analytical work to specific problems.

Parallel with these developments has been the development of analytical approximations for parametric inference in small samples, initiated by Fisher (1934) but largely overlooked until new developments were stimulated by Efron and Hinkley (1978) and Barndorff-Nielsen and Cox (1979). A flood of subsequent work is summarized in the books of Barndorff-Nielsen and Cox (1994), Pace and Salvan (1997), and Severini (2000). The efforts of many researchers, particularly O. E. Barndorff-Nielsen, (1983, 1986) and D. A. S. Fraser (e.g., Fraser (1990); Fraser, Reid and Wu (1999)) and their co-workers, have led to an elegant theory of near-exact inference based on small samples from parametric models. Its theoretical basis is saddlepoint and related approx-imation (Daniels, 1954, 1987), and further developments have been well described by Reid (1988, 1995, 2003). These methods are highly accurate in many situations, but are nevertheless under-used compared to the simulation procedures mentioned above. One reason for this may be their arcane basis in the conditionality principle, ancillary statistics and marginalization, and another may be the forbidding technical details, but the main reason is undoubtedly the lack of suitable software. Unlike the bootstrap libraries available in general-purpose languages such as S-PLUS (S-PLUS (2007)) and R (R Development Core Team (2007)) or specialized software such as WinBUGS (Lunn et al. (2000)) or *LogXact* (Cytel Inc. (2007)), small-sample parametric asymptotics have been implemented piecemeal, usually by specialists in the area for their personal use.

*Alessandra R. Brazzale is Associate Professor of Statistics, Department of Social, Cognitive and Quantitative Sciences (DSSCQ) University of Modena and Reggio Emilia, Viale Allegri 9, 42100 Reggio Emilia, Italy e-mail: alessandra.brazzale@unimore.it and Institute of Biomedical Engineering, Italian National Research Council. Anthony C. Davison is Professor of Statistics, Institute of Mathematics (IMA-FSB-EPFL), Ecole Polytechnique Fédérale de Lausanne, Station 8, 1015 Lausanne, Switzerland e-mail: anthony.davison@epfl.ch.*

This paper describes the implementation and use of libraries of software for higher order inference for several special classes of model: for linear exponential families such as logistic regression models, for nonnormal linear models and for nonlinear regression models with heteroscedastic normal errors. Its objective is to make higher order inference for such models available for use by those without a command of the technical details. We also describe how Markov chain Monte Carlo may be used not only to assess conditional coverage and related properties of some of our methods, but also for inference. A related, more extended, account may be found in Brazzale, Davison and Reid (2007), which gives many further examples. Butler (2007) gives ample evidence for the accuracy of the approximations that underlie some of the theory used herein.

Section 2 outlines developments in parametric asymptotics that undergird the numerical approximations whose implementation is described in Section 3. Application to logistic regression is described in Section 4, where we argue that the conservatism and fragility of exact inference in this context should lead us to prefer approximation. In Section 5 we discuss regression-scale models with nonnormal errors, outline how both analytical approximation and Markov chain Monte Carlo simulation may be used for approximate conditional inference, and compare them empirically. Section 6 describes how the approximate methods may be applied to nonlinear regression models, which are often fitted using small samples from bioassays or toxicological studies. The paper concludes with a brief discussion and appendices containing technical details.

## 2. BACKGROUND

### 2.1 First Order Inference

Initially we consider a parametric statistical model with density $f(y;\theta)$, where $\theta \in \Theta \subseteq \mathbb{R}^d$ is a $d$-dimensional parameter and $y = (y_1, \ldots, y_n)$ a vector of continuous responses. Let $L(\theta) \propto f(y;\theta)$ denote the likelihood and $\ell(\theta) = \log L(\theta)$ the log likelihood functions. Under mild conditions the maximum likelihood estimate $\hat\theta$ may be found by solving the score equation $\ell_\theta(\hat\theta) = 0$, and its asymptotic variance is approximated using the inverse of the observed information matrix $j(\hat\theta)$. We distinguish between quantities of primary interest and others not of direct concern by writing $\theta = (\psi, \lambda)$, where $\psi$ is a low-dimensional parameter of interest and $\lambda$ is a nuisance parameter whose dimension may be appreciably larger than that of $\psi$. This partitioning entails corresponding splits of the score vector $\ell_\theta(\psi,\lambda)$ into $\ell_\psi(\psi,\lambda)$ and $\ell_\lambda(\psi,\lambda)$, and of the observed information function $j(\psi,\lambda)$ into the sub-matrices $j_{\psi\psi}(\psi,\lambda)$, $j_{\psi\lambda}(\psi,\lambda)$, $j_{\lambda\psi}(\psi,\lambda)$ and $j_{\lambda\lambda}(\psi,\lambda)$.

Exact inference for linear exponential families and location-scale models was discussed by Fisher (1934) in a paper largely ignored for many years. Even where available, in principle, the effort needed to implement exact methods in all but the simplest cases means they are seldom used in practice, but are typically replaced by asymptotic approximations derived by assuming that the sample size $n$, or, more generally, some information index, tends to infinity. We then eliminate the nuisance parameter $\lambda$ by replacing it by the constrained maximum likelihood estimate $\hat\lambda_\psi$ obtained by maximizing $\ell(\psi,\lambda)$ with respect to $\lambda$ for fixed $\psi$. Inference about $\psi$ may then be performed using the profile log likelihood function $\ell_{\mathrm{p}}(\psi) = \max_\lambda \ell(\psi,\lambda) = \ell(\psi,\hat\lambda_\psi)$. The corresponding observed information function, $j_{\mathrm{p}}(\psi) = -\partial^2\ell_{\mathrm{p}}(\psi)/\partial\psi\,\partial\psi^{\mathsf{T}}$, can be expressed in terms of the full observed information function through the identity

$$j_{\mathrm{p}}(\psi) = j_{\psi\psi}(\hat\theta_\psi) - j_{\psi\lambda}(\hat\theta_\psi)\{j_{\lambda\lambda}(\hat\theta_\psi)\}^{-1} j_{\lambda\psi}(\hat\theta_\psi),$$

where $\hat\theta_\psi = (\psi, \hat\lambda_\psi)$. For scalar $\psi$, inference on the parameter of interest may be based on the Wald statistic, $j_{\mathrm{p}}(\hat\psi)^{1/2}(\hat\psi - \psi)$, score statistic, $\{j_{\mathrm{p}}(\hat\psi)\}^{-1/2}\ell_\psi(\psi,\hat\lambda_\psi)$, or on the likelihood root,

$$r(\psi) = \mathrm{sign}(\hat\psi - \psi)[2\{\ell_{\mathrm{p}}(\hat\psi) - \ell_{\mathrm{p}}(\psi)\}]^{1/2}, \quad (1)$$

which have standard normal distributions up to the order $O(n^{-1/2})$. When the sample size is small these first order approximations are often inaccurate, especially in complex models.

## 2.2 Higher Order Inference

The keys to refining the limiting behavior of the most important likelihood quantities are two higher order density approximations: Barndorff-Nielsen's (1983) $p^*$ formula and the tangent exponential model $p_{\mathrm{TEM}}$ developed by Fraser, Reid and Wu (1999). Apart from an $O(n^{-1})$ norming constant, the first gives the density of the maximum likelihood estimate at the observed data and at other points having the same value of an ancillary statistic, though these must be known. The second is an exponential model whose distribution function at the observed data differs from that of the conditional model by $O(n^{-3/2})$ under the observed conditioning (Fraser, Andrews and Wong (2005)). Both approximations are exact for transformation models and give excellent results generally. In full exponential families they agree with the saddlepoint approximation to the density of the minimal sufficient statistic given by Barndorff-Nielsen and Cox (1979).

These density approximations are mainly useful for deriving approximate distribution functions for appropriate pivots, from which we obtain P-values and confidence intervals for the parameters of interest. For scalar $\psi$, these approximate distribution functions have the forms

$$\Phi^*(r) = \Phi(r) + \phi(r)\left(\frac{1}{r} - \frac{1}{q}\right) \tag{2}$$

and

$$\Phi(r^*) = \Phi\left(r + \frac{1}{r}\log\frac{q}{r}\right), \tag{3}$$

for $r$ given at (1) and $q$ defined as

$$q = \frac{|\varphi(\hat\theta) - \varphi(\hat\theta_\psi)\ \varphi_\lambda(\hat\theta_\psi)|}{|\varphi_\theta(\hat\theta)|}\left\{\frac{|j(\hat\theta)|}{|j_{\lambda\lambda}(\hat\theta_\psi)|}\right\}^{1/2}, \tag{4}$$

where $\Phi(\cdot)$ and $\phi(\cdot)$ represent the standard normal distribution and density functions. Here, $\varphi(\theta)$ is a reparametrization based at the observed data and used to provide a third order distribution function approximation through the tangent exponential model, and $\varphi_\theta(\theta)$ and $\varphi_\lambda(\theta)$ represent the $d \times d$ matrix with $(i,j)$ element $\partial\varphi_i/\partial\theta_j$ and the $d \times (d-1)$ matrix with $(i,j)$ element $\partial\varphi_i/\partial\lambda_j$. Special expressions for (4) can be found in Appendix A.1. Expression (2) is known as a Lugannani–Rice-type approximation, and the quantity $r^*$ in the Barndorff-Nielsen-type approximation (3) is known as a modified likelihood root. Under ordinary repeated sampling, approximations (2) and (3) are exact up to the third order, that is,

$$\mathrm{pr}(R \le r; \theta) = \Phi^*(r) + O(n^{-3/2}),$$

$$\mathrm{pr}(R^* \le r^*; \theta) = \Phi(r^*) + O(n^{-3/2}).$$

In comparison, the likelihood root $r$ itself is standard normal only to the first order, $O(n^{-1/2})$. A rather subtle Taylor series expansion of $\Phi(r^*)$ around $r$, taking into account the dependence of $\varphi(\theta)$ on the observed data point, shows that $\Phi(r^*) = \Phi^*(r) + O(n^{-3/2})$, rising to $O(n^{-1})$ if this dependence is not accommodated; in particular the more accurate results holds for linear exponential families (Jensen (1992)), in which $\varphi(\theta)$ does not depend on the observed data. In an exponential family of order one, $\Phi^*(r)$ equals the Lugannani and Rice (1980) tail area approximation. In the presence of nuisance parameters, it gives the approximation due to Skovgaard (1987).

## 2.3 Related Ideas

An alternative to the ideas outlined in Section 2.2 is first to adjust the profile likelihood $L_{\mathrm{p}}(\psi) = \exp\{\ell_{\mathrm{p}}(\psi)\}$ to account for the presence of nuisance parameters, and then to correct the first order statistics obtained therefrom in order to improve the standard normal approximation. Pierce and Peters (1992) call these sequential approximations, as contrasted with the more common double approximations (Barndorff-Nielsen and Cox (1979)).

The general form of the adjusted profile likelihood is

$$L_{\mathrm{a}}(\psi) = L_{\mathrm{p}}(\psi)M(\psi), \tag{5}$$

with suitably defined correction term $M(\psi)$; see Appendix A.2. When an exact conditional or marginal likelihood function for $\psi$ exists, this is approximated to the order $O(n^{-1})$ by the adjusted profile likelihood function. In stratified models with the number of nuisance parameters proportional to the number of strata, Sartori et al. (1999) showed that $M(\psi)$ corrects for the presence of the nuisance parameters. The maximizing value $\hat\psi_{\mathrm{a}}$ usually has a smaller finite-sample bias than does $\hat\psi$, and the likelihood root $r_{\mathrm{a}}(\psi)$ based on $L_{\mathrm{a}}$ has a distribution that is closer to normal than does $r(\psi)$. Insight into why likelihood roots obtained from adjusted likelihoods tend to achieve most of the distributional improvement achieved by higher order methods, despite their null distribution being standard normal only to the first order, is provided by Sartori (2003) in a two-index asymptotics setting and by DiCiccio and Efron (1996), who relate their findings to the bootstrap.

If the parameter of interest is vector, formula (3) cannot be used. Skovgaard (2001) suggests adjusting the likelihood ratio statistic $w(\psi) = 2\{\ell_p(\hat{\psi}) - \ell_p(\psi)\}$, which has the $\chi^2$ distribution with $d_0 = \dim(\psi)$ degrees of freedom up to the order $O(n^{-1})$. His proposed adjusted likelihood ratio statistic

$$w^* = w\{1 - w^{-1} \log u\}^2, \tag{6}$$

with correction term $u$ suitably defined, is also asymptotically distributed as $\chi^2_{d_0}$, but behaves much better than $w$ in small samples. If $\psi$ is scalar, then $u = r/q$ with $q$ given by (4), and (6) reduces to $(r^*)^2$.

The above discussion applies only to continuous response models. For discrete responses, analogous results are in general unavailable. However, for distributions such as the binomial and Poisson whose support has, or can easily be transformed to, a lattice structure, the use of a slightly modified form of (2) provides approximations to tail probabilities with error $O(n^{-1})$ (Severini (2000), Section 6.3.3). Pierce and Peters (1992) discuss continuity correction for the asymptotic approximations. As discussed in Section 4, the uncorrected version can be interpreted as an approximation to the mid-P value (Agresti (2002), page 20)

$$p_{\text{mid}}(x; \psi) = \text{pr}(X < x; \psi) + \tfrac{1}{2}\,\text{pr}(X = x; \psi)$$

for a suitable lattice random variable $X$. Further discussion is given by Pierce and Peters (1999), Davison and Wang (2002), and Davison, Fraser and Reid (2006), who indicate that use of $r^*$ unmodified in standard discrete cases approximates the mid-P value with error of the order $O(n^{-1})$. From a practical point of view, the most reassuring point is perhaps not the precise asymptotic order of these approximations, but rather the fact that they are relative, and thus give accurate values even for small tail probabilities.

## 3. IMPLEMENTATION

### 3.1 General

Many journal pages and several books have been devoted to the ideas sketched in Section 2, but their widespread adoption in practice has been limited by the lack of suitable software. The `R` package bundle `hoa`, short for higher order asymptotics, is intended to make these methods readily accessible by providing easy-to-use and self-contained code for routine data analysis with logistic regression models, nonnormal linear models and nonlinear models with nonconstant variance (Brazzale, 2005). These models are widely used in applications: logistic regression is a common tool in epidemiology and medicine; nonnormal linear models comprise models used in survival and reliability analyses; and nonlinear heteroscedastic models are increasingly used in biostatistics, for instance, in herbicide bioassays and ecotoxicity tests. The code is organized as four packages, three of which—`cond`, `marg` and `nlreg`—refer to the model classes just mentioned. A fourth—`csampling`—contains conditional sampling routines for nonnormal linear models and was used to produce the results presented in Section 5. The code is freely available from `http://statwww.epfl.ch/AA` or can be downloaded from CRAN (http://cran.r-project.org).

The remainder of this section sketches the core ideas that make it possible to implement higher order asymptotics in a numerical computing environment with limited facilities for algebraic manipulation. The issues inherent to the implementation for logistic and nonlinear models are described in Brazzale (1999) and Bellio and Brazzale (2003), respectively. The complete design strategy can be found in Brazzale (2000), Chapter 6.

### 3.2 Building-Blocks

The complexity of the algebraic expressions involved is the main obstacle to the implementation of higher order asymptotics in numerical computing environments, most of which have inefficient symbolic manipulation capabilities, if any at all. The key to our implementation of the methods presented in Section 2 is to identify building-blocks into which the higher order statistics can be decomposed and which are provided or can efficiently be handled by the computing device. This builds upon the observation of Davison (1988) that in linear exponential families the output of standard fitting routines suffices to calculate the $q(\psi)$ and $M(\psi)$ correction terms of Section 2. Brazzale (2000), Section 6.1, derives the corresponding building-blocks for nonnormal linear and nonlinear heteroscedastic regression models. The first of these classes is characterized by the design matrix $X$, the standardized residuals $a$, minus the logarithm of the density function of the error term, $g_0(\varepsilon) = -\log f(\varepsilon)$, and its first two derivatives (see Brazzale, Davison and Reid (2007), Section 8.6.2). For nonlinear regression models the only requirements are the mean and variance functions $\mu(x;\beta)$ and $w(x;\beta,\rho)^2$ and their first two derivatives (see Brazzale, Davison and Reid (2007), Section 8.6.3). Two design strategies were adopted to make these quantities available in `hoa`: either they are provided by special constructs, called *family objects*, or they are derived as needed by exploiting the algebraic manipulation function `deriv3` available in `R`; see Bellio and Brazzale (2003).

### 3.3 Pivot Profiling

Inferences provided by fitting routines available in statistical computing environments such as S-PLUS and `R` are generally based upon first order asymptotics. Most often the Wald statistic is used, because its linearity in the interest parameter $\psi$ yields simple readily computed confidence intervals. The likelihood ratio statistic is parametrization-invariant, and hence more reliable, but its nonlinearity in $\psi$ means that construction of confidence intervals entails re-fitting the model for all the required values of $\psi$. We deal with this by using cubic splines to interpolate values of $r(\psi)$ and related quantities among values calculated exactly for a grid encompassing the required range of $\psi$. In `R` this may be achieved by using different *offsets*, for each of which the necessary output is retrieved and statistics calculated. Estimates and confidence bounds are read off from significance functions such as (2) and (3) using the fitted splines.

Numerical interpolation of higher order solutions works very well for analytic functions such as the profile and adjusted profile likelihoods, but quantities such as $r^*$ have a singularity at $\psi = \hat{\psi}$, and the numerical values calculated may be unstable if $\psi$ is close to $\hat{\psi}$. This problem is particularly acute for logistic regression. Nonnormal linear models are much less affected, and in our experience numerical instabilities are almost absent for nonlinear models. To avoid singularities in the first two model classes, we implemented a hybrid algorithm that uses two-step polynomial interpolation of the higher order statistics for values of $\psi$ in a small interval around $\hat{\psi}$. The higher order solutions are expressed as polynomials of the likelihood root $r$, which itself is written as a polynomial in $\psi$, and the coefficients of these polynomials are estimated by least squares. For nonlinear models, it suffices to avoid exact computation of the higher order statistics for values of $\psi$ very close to the maximum likelihood estimate.

The procedure just described represents the bulk of the approximate conditional inference routines in the `cond`, `marg` and `nlreg` packages, which enable inference for the three model classes of the `hoa` bundle. See Section 6.3.2 and Appendix B.2 of Brazzale (2000) for further details.

### 3.4 Markov Chain Monte Carlo

The generation of observations conditional on an ancillary statistic is useful for inferential purposes such as the calculation of confidence intervals and P-values whenever the exact conditional density is unknown or difficult to obtain without simulation. Such an approach is described in an unpublished technical report by Casella, Wells and Tanner (1994), who emphasize sampling-based calculations for pivotal inference using the Gibbs sampler.

Conditional inference may also be used to assess the quality of small-sample solutions. Studies of the properties of the methods presented in Section 2 (DiCiccio, Field and Fraser (1990), DiCiccio and Field (1991), Ronchetti and Ventura (2000)) focused on their numerical accuracy, stability and sensitivity to model failure. So far as we know, there has been no numerical investigation of these properties conditional on an ancillary statistic; Severini (1999) and Ventura (1997) grouped their simulation results by the levels of two nearly independent functions of the ancillary, but this does not amount to a fully conditional simulation. Trotter and Tukey (1956) were the first to simulate conditionally on an ancillary statistic in the special case of normal samples, but there have been few contributions since (Durbin (1961); Bondesson (1982); Fraser, Lee and Reid (1990); Morgenthaler and Tukey (1991)).

The `csampling` package of the `hoa` bundle includes a conditional sampler for general regression-scale models and extends Bondesson's (1982) method by replacing an acceptance-rejection algorithm by the Metropolis–Hastings algorithm (Robert and Casella (2004), Chapter 7). Section 5.2 describes a simulation study performed using this package.

## 4. LOGISTIC REGRESSION

### 4.1 Likelihood Approximation

After the linear model, logistic regression of binary responses $y_1,\ldots,y_n$ on covariates $(z_1,x_1),\ldots,$

$(z_n, x_n)$ is perhaps the most widely used parametric regression procedure. Let $X$, $Z$ and $y$ denote the $n \times p$, $n \times k$ and $n \times 1$ matrices whose $i$th rows are respectively $x_i^\top$, $z_i^\top$ and $y_i$. The log likelihood

$$\ell(\psi, \lambda) = y^\top Z\psi + y^\top X\lambda - \sum_{i=1}^{n} \log\{1 + \exp(z_i^\top \psi + x_i^\top \lambda)\}$$

corresponds to a linear exponential family with canonical parameter $(\psi, \lambda)$ and sufficient statistic $(t, s) = (Z^\top y, X^\top y)$, so the higher order approximations described above are particularly simple and can be obtained by re-arranging the output of standard routines for fitting logistic models (Davison (1988)); see also Daniels (1958), Barndorff-Nielsen and Cox (1979), Pierce and Peters (1992) and Strawderman, Casella and Wells (1996). Such approximations are provided by our `cond` package.

Example 1 (*Urine data*). For illustration, we take data on the presence or not of crystals in urine samples (Andrews and Herzberg (1985), page 249). Full data on the six quantitative covariates are available for 77 individuals, and we consider the coefficient $\psi$ of the variable `urea` representing urea concentration (millimoles/litre) in a logistic regression model also containing the five other covariates and an intercept. The `R` code for first order and higher order inferences is

```
> uri.glm <- glm( r ~ gravity + ph +
+                     osmo + conduct +
+                     urea + calc,
+                 family = binomial,
+                 data = urine )
> summary( uri.glm )
> uri.urea <- cond( uri.glm,
+                   offset = urea )
> summary( uri.urea )
> plot( uri.urea )
```

The first two instructions fit the model by maximum likelihood and then print the results, and the last four lines of code compute, summarize and plot the first and higher order approximations. The maximum likelihood estimate and its standard error are $\hat{\psi} = -0.0320$ (0.0161), yielding a Wald statistic of $-1.99$ with two-sided P-value $2\Phi(-1.99) = 0.047$. An approximation to the conditional maximum likelihood estimate and its standard error is obtained by maximizing the adjusted profile log likelihood and taking its curvature at the maximum; this gives $\hat{\psi}_{\rm a} = -0.0276$ (0.0149), yielding an approximate conditional Wald statistic of $-1.85$ and P-value 0.064. The 95% confidence intervals for $\psi$ based on these two Wald pivots are respectively $(-0.0636, -0.0004)$ and $(-0.0568, 0.0016)$, and those based on the likelihood root $r$, and the modified likelihood root $r^*$ are $(-0.0668, -0.0025)$ and $(-0.0587, 0.0005)$. Thus, the estimated coefficient changes by around 14%, more than might be anticipated with six nuisance parameters and 77 observations, and there are corresponding changes to the confidence intervals.

Figure 1 shows two of the graphs provided by the `plot` command: note the large difference between first and higher order inference summaries, which suggests that the latter should be used as a matter of course with binary data models. The adjusted profile likelihood corrects for the finite sample bias in the maximum likelihood estimator, in analogy to the conditional likelihood function, while the modified likelihood root also accounts for the nonnormality of the ordinary and adjusted likelihood functions.

A simple information computation sheds some light on the size of the higher order correction in the example above. Suppose that we have independent observations $y_1', \ldots, y_n'$ from the logistic density $\exp(y_i' - \lambda - z_i\psi)/\{1 + \exp(y_i' - \lambda - z_i\psi)\}^2$, where the $z_i$ are known scalar covariates. The asymptotic variance $v_{\rm cont}$ of the maximum likelihood estimator of $\psi$ based on the continuous $y_i'$ is a corner of the Fisher information matrix, which is easily seen to be $3(\tilde{X}^\top \tilde{X})^{-1}$, where $\tilde{X}$ denotes the entire design matrix, whose $i$th row here is $(1, z_i)$. If only the sign of the $y_i'$ is known, so the continuous observations are replaced by binary variables, the asymptotic variance of the maximum likelihood estimator of $\psi$ is a corner of the matrix $(\tilde{X}^\top W \tilde{X})^{-1}$, where $W$ denotes the $n \times n$ diagonal matrix $\mathrm{diag}\{\pi_1(1-\pi_1), \ldots, \pi_n(1-\pi_n)\}$, with $\pi_i = \exp(\lambda + z_i\psi)/\{1 + \exp(\lambda + z_i\psi)\}^2$. The ratio of these asymptotic variances gives some idea of the information content of the binary data compared to the continuous data. Figure 2 shows this ratio as a function of the standardized parameter $\delta_{\rm cont} = \psi/(v_{\rm cont})^{1/2}$ when the covariate $z$ takes $n = 21$ equi-spaced values ranging from $-3$ to 3. The maximum value 0.75 occurs when $\lambda = \psi = 0$, but the ratio drops fast as $\delta_{\rm cont}$ increases. A value of $\delta_{\rm cont} = 5$ that would be easily distinguished from 0 using the continuous data would correspond to a value of around 2 based on the binary data, and this would be much less easily distinguished from zero.

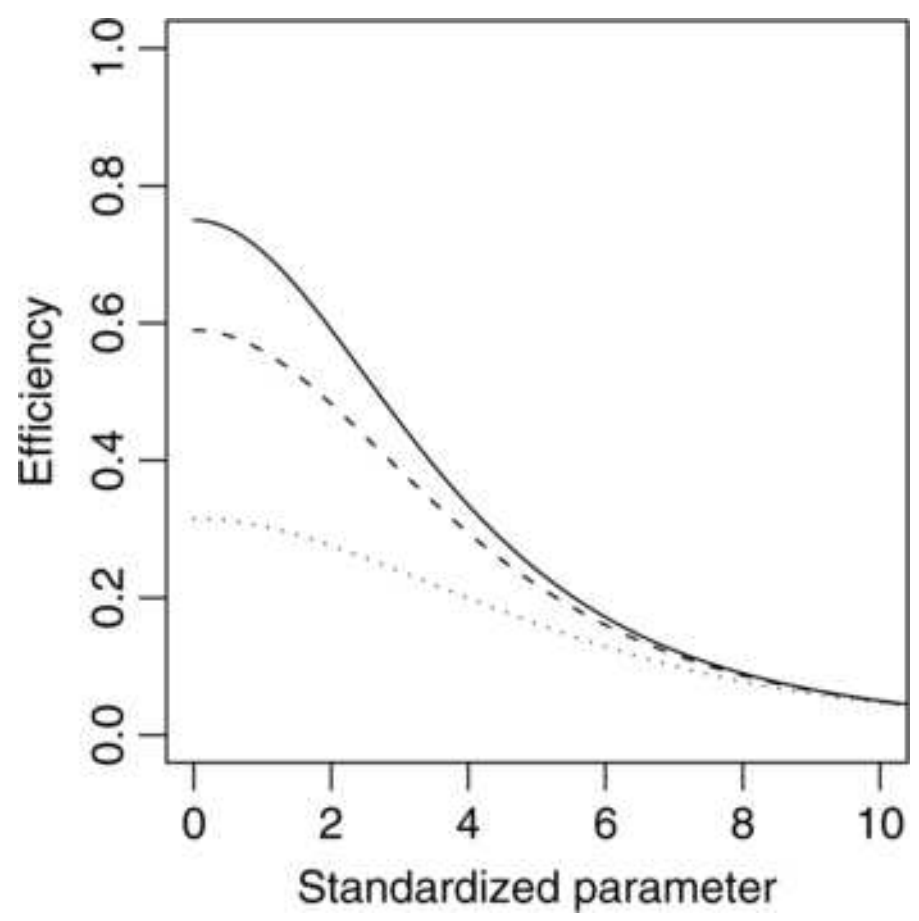

FIG. 2. *Efficiency of logistic regression for estimation of $\psi$ relative to estimation with corresponding continuous responses from the logistic distribution, as a function of the standardized parameter $\delta$ for the continuous model, for $\lambda = 0$ (solid), $\lambda = 1$ (dashes) and $\lambda = 2$ (dots).*

If the same computation is applied to the urine data, then the numbers of continuous observations equivalent to the 77 binary observations range from 7 to 19, depending on the parameter considered, with a value of 16 or so for $\psi$. In this light the difference between first order and higher order results seems much more explicable: we are fitting a model with 6 nuisance parameters to the equivalent of fewer than 20 continuous observations, and so one would expect an appreciable "degrees of freedom" adjustment. Section 4.2 of Brazzale, Davison and Reid (2007) gives related discussion.

## 4.2 Exact Inference

In a logistic regression model, exact inference for the interest parameter $\psi$ is available from the conditional density function of $T$ given the value of $S$,

$$\text{pr}(T = t \mid S = s; \psi) = \frac{\exp(y^\top Z\psi)}{\sum_{u \in \mathcal{A}_s} \exp(u^\top Z\psi)}, \tag{7}$$

where $\mathcal{A}_s = \{y : y^\top X = s, y \in \{0,1\}^n\}$. The function (7) does not depend on $\lambda$, and this greatly simplifies inference (Cox (1958)). The main practical difficulty in using (7) is the enumeration of the elements of $\mathcal{A}_s$, but recent computational advances have brought this into reach, at least in simple cases. One possibility is to use the network algorithm (Mehta and Patel (1995), Mehta, Patel and Senchaudhuri (2000)) provided by commercial software such as `LogXact`, but although helpful in simple problems, it can be impractically slow when there are several covariates. Forster, McDonald and Smith (2003) propose a Markov chain Monte Carlo algorithm for more complex models, but as their chain may be reducible, there is no guarantee that $\mathcal{A}_s$ would be fully explored even if the chain were to be run forever. Their algorithm has been implemented in the `elrm` package of `R` by Zamar, McNeney and Graham (2007).

Apart from the enumeration of $\mathcal{A}_s$, there are two deeper problems, both linked to the exactness of (7): the conservatism of exact tests and confidence intervals, which leads to overly-wide intervals and overly-large P-values, and the fragility of exact conditional inference in certain discrete cases. We now discuss these, illustrated by data with $n=16$ binary responses:

$$y^{\mathsf{T}} = (1\ 0\ 1\ 0\ 1\ 1\ 1\ 0\ 0\ 0\ 0\ 0\ 1\ 0\ 0\ 0),$$

covariate matrices

$$X^{\mathsf{T}} = \frac{1}{2}\begin{pmatrix} 2 & 2 & 2 & 2 & 2 & 2 & 2 & 2 & 2 & 2 & 2 & 2 & 2 & 2 & 2 & 2 \\ -3 & -3 & -3 & -3 & -1 & -1 & -1 & -1 & 1 & 1 & 1 & 1 & 3 & 3 & 3 & 3 \end{pmatrix},$$

$$Z^{\mathsf{T}} = \frac{1}{2}(-3\ -1\ 1\ 3\ -3\ -1\ 1\ 3\ -3\ -1\ 1\ 3\ -3\ -1\ 1\ 3),$$

and corresponding sufficient statistics $s = X^{\mathsf{T}}y = (6,-3)$ and $t = Z^{\mathsf{T}}y = -4$. In this case the full sample space has $2^{16}$ elements, reducing to 8008 and 13 elements respectively when conditioned on the first component of $s$ and on both components of $s$. We use this example in the next two subsections.

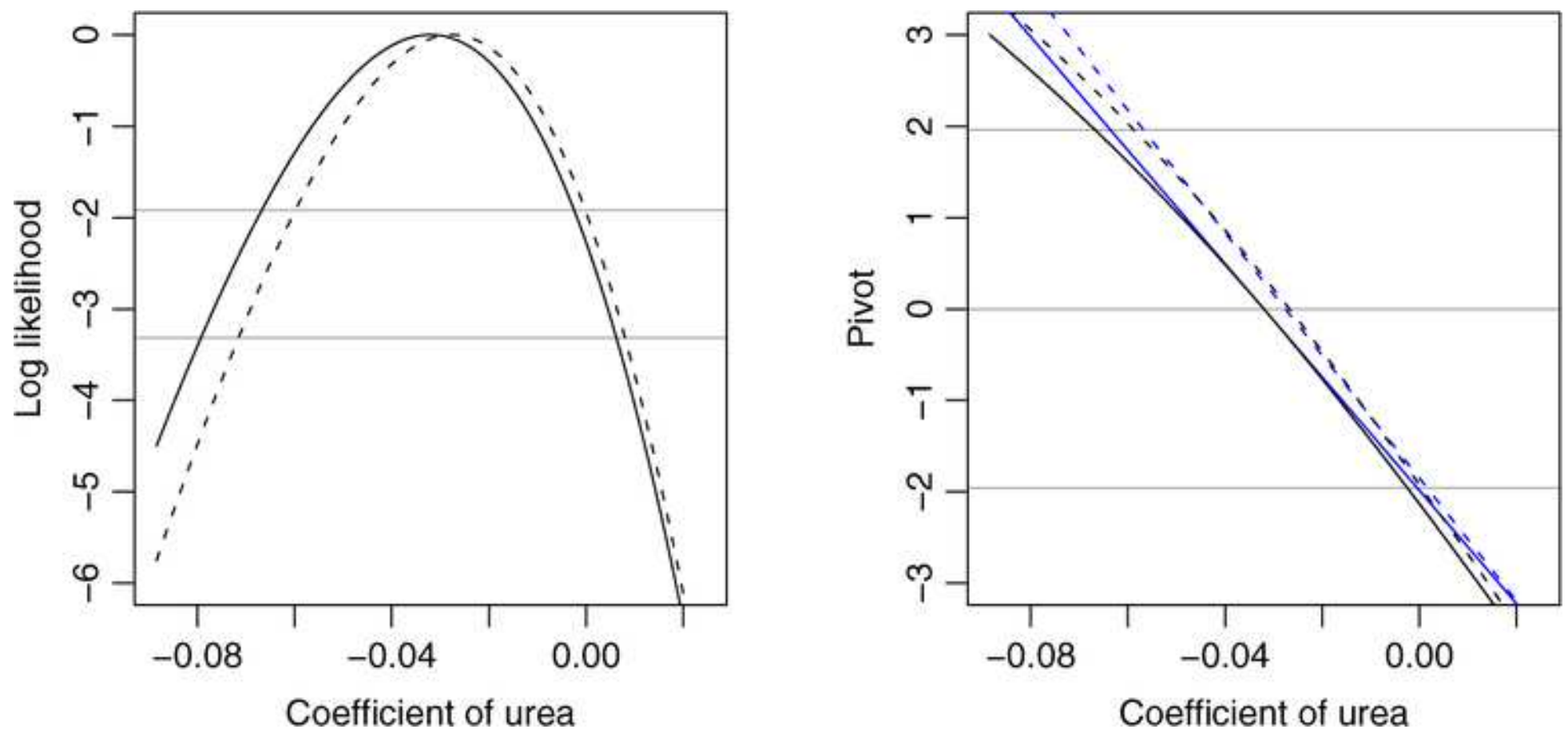

FIG. 1. *Graphical output comparing first (solid) and higher order (dashes) inference summaries for the urine data. Left: profile log likelihood $\ell_p(\psi)$ (solid) and adjusted profile log likelihood $\ell_a(\psi)$ (dashes), with horizontal gray lines indicating 0.95 and 0.99 confidence sets for $\psi$. Right: likelihood root $r(\psi)$ (solid, curved) and modified likelihood root $r^*(\psi)$ (dashes, curved), and Wald pivots based on the profile likelihood (solid, straight) and the adjusted profile likelihood (dashes, straight). The horizontal gray lines are at the* 0.025, 0.5 *and* 0.975 *standard normal quantiles.*

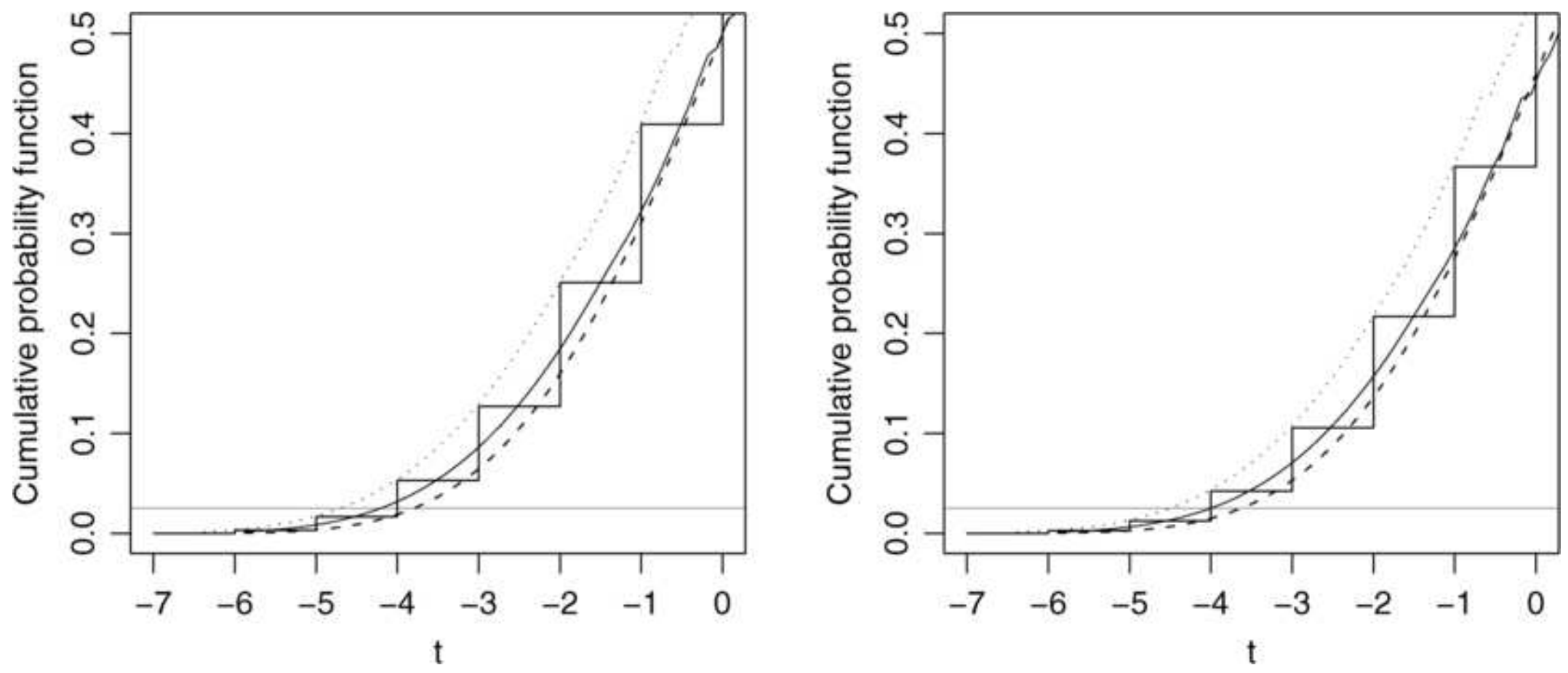

Fig. 3. *Exact conditional distribution of $T$ in logistic example (step function), with $\Phi\{r^*(t)\}$ (solid), $\Phi\{r(t)\}$ (dashes), and $\Phi\{r^*(t+1/2)\}$ (dots), shown for $t=-6,\dots,0$. Left: $\psi=0$. Right: $\psi=0.05$. The horizontal gray lines are at 0.025.*

### 4.3 Conservatism of Exact Inference

Exact inference in discrete response models typically leads to conservative tests and confidence intervals. A striking illustration of this in the context of simple binomial models is given by Agresti and Coull (1998), who show how exact intervals such as that due to Clopper and Pearson (1934) are conservative for all values of the underlying parameter, while approximate intervals based on likelihood quantities have overall coverage closer to the nominal level. For a variety of viewpoints on this and some solutions, see Agresti and Caffo (2000), Brown, Cai and DasGupta (2001) and Geyer and Meeden (2005).

In the case of our simple example, Figure 3 shows the lower tail of the exact conditional distribution function of $T$ when $\psi=0$, obtained by computing the generating function for the combinatorial terms; it has support on the set $\{-6,-5,\dots,6\}$. Also shown are the approximate conditional distributions obtained by taking $\Phi\{r(t;0)\}$, $\Phi\{r^*(t;0)\}$, and $\Phi\{r^*(t+1/2;0)\}$, for a grid of values of $t$ in the range $(-6,6)$; the corresponding datasets were constructed to retain the original value of the conditioning statistic $s$. These approximations correspond respectively to first order and higher order procedures, and to use of the higher order procedure with a continuity correction. Use of the function $\Phi\{r^*(t;\psi)\}$ for $\psi=0$ yields a continuous approximation to the exact conditional distribution function that closely matches the mid-points of the jumps in the step-function and thus the mid-P value.

Table 1 compares P-values and confidence limits for these data. The results for mid-P and the modified likelihood root are fairly close, and give tighter inferences than does the exact solution, which is close to the modified likelihood root, plus continuity correction. The kink at $t=0$ in the approximations involving $r^*$ is due to a numerical instability. Although a different expression given as a limit for $r\to 0$ is available, it is rarely used in practice because errors in P-values that are close to 0.5 are unimportant.

TABLE 1

*One-sided P-values for testing $\psi = 0$ and limits of nominally equi-tailed 95% confidence intervals for $\psi$, for the artificial logistic regression example*

| | **P-values** | **Limits of confidence interval** |
|---|---|---|
| Exact | 0.0528 | (−2.992, 0.158) |
| mid-P | 0.0346 | (−2.690, 0.069) |
| Wald | 0.0399 | (−2.572, 0.144) |
| Wald, modified | 0.0475 | (−2.290, 0.183) |
| Likelihood root, $r$ | 0.0190 | (−2.950, −0.060) |
| Modified likelihood root, $r^*$ | 0.0318 | (−2.506, 0.050) |
| with continuity correction | 0.0557 | (−2.906, 0.172) |

The equi-tailed exact confidence interval $(\psi_-, \psi_+)$ with level $(1 - 2\alpha)$ has limits given by the solutions to the equations

$$\text{(8)} \qquad \begin{aligned} \text{pr}(T \geq t \mid S = s; \psi_-) &= \alpha, \\ \text{pr}(T \leq t \mid S = s; \psi_+) &= \alpha, \end{aligned}$$

whereas the limits of the intervals based on $r$ and $r^*$ are the solutions in $\psi$ of the equations

$$\Phi\{r(t;\psi)\} = \alpha, 1 - \alpha, \quad \Phi\{r^*(t;\psi)\} = \alpha, 1 - \alpha,$$

respectively. The right-hand panel of Figure 3 shows the conditional distribution for $T$ for $\psi = 0.05$, which for $t < 0$ slightly depresses the probabilities relative to taking $\psi = 0$, and illustrates why the exact intervals are wider, and hence more conservative, than are these approximate ones: it is necessary to take $\psi_+ > 0.05$ to satisfy the right-hand equation in (8); in fact, the first line of Table 1 shows that $\psi_+ = 0.158$ is required.

## 4.4 Fragility of Exact Conditional Inference

The second issue is the sensitivity of the set $\mathcal{A}_s$, and hence of exact conditional inference, to the matrix $X$. It seems reasonable to require that small changes to $X$, for instance, due to rounding of the explanatory variables, should lead to small changes in confidence intervals and P-values. To test this, we jittered the second column $x_2$ of the matrix $X$ in the simple example of Section 4.2. When the elements of $x_2$ were perturbed by adding normal noise with standard deviation 0.01, rounded to 3 decimal places, the value of $s$ changed to $(6, -3.013)$, and the support points of the conditional distribution reduced from $\{-6, -5, \ldots, 6\}$ to $\{-4, -2, 0\}$. The exact tail probability for $t$, $\text{pr}(T \leq t \mid S = s; \psi = 0)$, changed from 0.0528 to 0.3333, and mid-P from 0.0347 to 0.167, but $\Phi\{r^*(t)\}$ changed only from 0.0318 to 0.0316.

An attempt to assess the fragility of the inference for the urine data failed: when the covariates are scaled to zero mean and unit variance, and rounded to the nearest integer, giving 5–6 rounded values for each covariate, one million iterations of the algorithm described by Forster, McDonald and Smith (2003), designed to enumerate the conditional sample space for the urea effect, found 13 support points. A Markov chain run with rounding to the first decimal place failed to move at all, suggesting that, with this degree of precision in the covariates, the conditional distribution for the urea effect is degenerate. Thus, exact conditional inference seems to be out of reach for these data.

To compare more systematically the effects of perturbing the covariate on exact and approximate conditional inferences, we repeated this experiment 1000 times, by adding noise with standard deviation 0.01 to $x_2$, and rounding to different precisions. Table 2 gives the sizes of the resulting conditional sample spaces. Small changes to the covariates may sharply change the conditional sample space, and thus may severely affect exact conditional inferences and derived quantities such as mid-P values, but the approximate conditional inferences barely change: for each level of rounding, the average values of $r$ and $r^*$ were −2.076 and −1.855 across the 1000 simulated datasets, with standard errors of around 0.004 and 0.0035 for all levels of rounding; the values of $r$ and $r^*$ are of course constant when rounding to one decimal place. The mid-P values are computed with respect to the exact distribution, and hence are very sensitive to changes in the covariates; the 'approximating mid-P value' $\Phi(r^*)$ might in such cases be

TABLE 2

*Changes in number of support points of conditional sample space for a logistic regression model when a covariate is perturbed by adding small amounts of noise, rounded to different numbers of decimal places. As the number of decimal places increases, the conditioning becomes increasingly restrictive*

| **Decimal places** | **Number of support points of conditional sample space** | | | | | | | | | | | | |
|---|---|---|---|---|---|---|---|---|---|---|---|---|---|
| | **1** | **2** | **3** | **4** | **5** | **6** | **7** | **8** | **9** | **10** | **11** | **12** | **13** |
| 1 | | | | | | | | | | | | | 1000 |
| 2 | 9 | 8 | 16 | 13 | 33 | 50 | 65 | 103 | 150 | 206 | 181 | 138 | 28 |
| 3 | 129 | 212 | 191 | 167 | 141 | 101 | 38 | 15 | 4 | 1 | 1 | | |
| 4 | 730 | 221 | 46 | 3 | | | | | | | | | |

regarded as having been computed from an approximating continuous distribution (Davison and Wang (2002)).

These results supplement the finding of Pace, Salvan and Ventura (2004) that rounding of the response has little effect on higher order likelihood procedures.

## 5. REGRESSION-SCALE MODELS

### 5.1 Exact Inference

Nonnormal linear models, also known as regression-scale models, belong to the wider class of transformation models. They may be written using matrix notation as

$$y = X\beta + \sigma\varepsilon, \tag{9}$$

where $X$ is a fixed $n \times p$ design matrix with unknown regression coefficient $\beta \in \mathbb{R}^p$, $\sigma > 0$ is a scale parameter, and $\varepsilon = (\varepsilon_1, \ldots, \varepsilon_n)$ represents an $n$-dimensional vector of errors which are independently and identically distributed according to a known though not necessarily normal density $f(\cdot)$ on $\mathbb{R}$. If the maximum likelihood estimates $(\hat{\beta}, \hat{\sigma})$ exist and are finite, there exists a one-to-one change of variables from $y = (y_1, \ldots, y_n)$ to $(\hat{\beta}, \hat{\sigma}, a)$, where $a_i = (y_i - x_i^{\mathsf{T}}\hat{\beta})/\hat{\sigma}$, $i = 1, \ldots, n$, are the standardized residuals of the model and $x_i^{\mathsf{T}}$ is the $i$th row of $X$. The pair $(\hat{\beta}, \hat{\sigma})$ forms a transformation variable, whereas the vector of standardized residuals $a = (a_1, \ldots, a_n)$ is ancillary with respect to $\beta$ and $\sigma$. As shown by Fraser (1979), Section 6.1.5, the functionally unique separation of $\beta$ and $\sigma$ is obtained from the pivots $Z_1 = (\hat{\beta} - \beta)/\hat{\sigma}$ and $Z_2 = \hat{\sigma}/\sigma$, whose joint distribution given $a$ is known up to a normalizing constant. Fisher (1934), Fraser (1979) and others suggest that inference on the parameters $(\beta, \sigma)$ should be made conditionally on $a$. Conditional confidence intervals for single parameters are based on the marginal densities of the corresponding pivots, obtained by integrating out the remaining components in

$$\begin{aligned} &f_{Z_1,Z_2|a}(z_1, z_2 \mid a) \\ &\quad = c(a) z_2^{n-1} \prod_{i=1}^{n} f\{(x_i^{\mathsf{T}} z_1 + a_i) z_2\} |X^{\mathsf{T}} X|^{1/2}. \end{aligned} \tag{10}$$

Lawless (1972, 1973, 1978) gives applications to Cauchy, logistic, Weibull, and extreme value distributions and, Kappenmann (1975) to the Laplace distribution under a location-scale model.

Exact calculation of the marginal distribution for the pivots of interest usually involves multidimensional numerical integration, and can be difficult. For instance, the normalizing constant is given by

$$\begin{aligned} c(a)^{-1} = \int_0^{+\infty} \int_{-\infty}^{+\infty} \cdots \int_{-\infty}^{+\infty} & z_2^{n-1} \\ & \cdot \prod_{i=1}^{n} f\{(x_i^{\mathsf{T}} z_1 + a_i) z_2\} \\ & \cdot |X^{\mathsf{T}} X|^{1/2} \, dz_{11} \\ & \cdots dz_{1p} \, dz_2, \end{aligned}$$

where $z_{1l} = (\hat{\beta}_l - \beta_l)/\hat{\sigma}$. The required computational effort rapidly becomes infeasible, especially if the number of parameters is large and the dimension of the interest parameter is low. There are two ways to overcome this problem. The first is to use the higher order theory presented in Section 2, which applies rather naturally to regression-scale models. The tail area approximations (2) and (3) agree with those proposed by DiCiccio, Field and Fraser (1990) and by Fraser, Lee and Reid (1990) for the marginal distribution functions of the pivots $Z_1$ and $Z_2$. All these methods are available through the package `marg` of the `hoa` bundle, which is equivalent in its design, syntax and use to the `cond` package. Examples of application are given in Section 5.2 of Brazzale, Davison and Reid (2007) and in Section 5.3.2 of Brazzale (2000).

The second way to avoid numerical calculation of the normalizing constant in (10) is to use Markov chain Monte Carlo (MCMC) techniques to simulate from the conditional distribution.

### 5.2 Monte Carlo Simulation

Classical simulation techniques generate observations that are independent and identically distributed by sampling directly from the target density. In our case this is not possible, as the normalizing constant $c(a)$ in (10) is generally unknown. Among possibilities for dealing with this are the conditional sampler available through the `rsm.sample` routine of the `csampling` package, which implements the Metropolis–Hastings algorithm. This routine samples not from the conditional density (10) of the pivots, but from that of the maximum likelihood estimates $(\hat{\beta}, \hat{\sigma})$, namely,

$$f_{\hat{\beta},\hat{\sigma}|a}(\hat{\beta}, \hat{\sigma} \mid a; \beta, \sigma)$$

$$(11)\qquad = c(a)\frac{\hat{\sigma}^{n-p-1}}{\sigma^n}\prod_{i=1}^{n} f[\{x_i^{\mathsf{T}}(\hat{\beta}-\beta)+\hat{\sigma}a_i\}/\sigma] \cdot |X^{\mathsf{T}}X|^{1/2}.$$

Because of the one-to-one relationship between the maximum likelihood estimates and the pivots $(Z_1, Z_2)$ given $a$, both approaches yield the same results, but sampling from (11) makes it easier to investigate the distributions of higher order statistics. The pseudocode for the conditional sampler may be written as:

- `Choose a candidate generation density` $f_c(\cdot)$.
- `Choose an initial value` $(\hat{\beta}_0, \hat{\sigma}_0)$.
- `For` $t = 1, \ldots, T$

  1. `Generate` $(\hat{\beta}_c, \hat{\sigma}_c)$ `from` $f_c(\cdot)$. `Take`

  $$(\hat{\beta}_t, \hat{\sigma}_t) = \begin{cases} (\hat{\beta}_c, \hat{\sigma}_c) \\ \quad \texttt{with probability}\ \pi, \\ (\hat{\beta}_{t-1}, \hat{\sigma}_{t-1}) \\ \quad \texttt{with probability}\ 1-\pi, \end{cases}$$

  `where`

  $$\pi = \min\left\{\frac{f(\hat{\beta}_c, \hat{\sigma}_c \mid a; \beta_0, \sigma_0) f_c(\hat{\beta}_{t-1}, \hat{\sigma}_{t-1})}{f(\hat{\beta}_{t-1}, \hat{\sigma}_{t-1} \mid a; \beta_0, \sigma_0) f_c(\hat{\beta}_c, \hat{\sigma}_c)},\ 1\right\},$$

  $a$ `is the ancillary and` $(\beta_0, \sigma_0)$ `the simulation parameters.`
  2. `Reconstitute the sample` $y_t = (y_{1t}, \ldots, y_{nt})$, `where` $y_{it} = x_i^{\mathsf{T}}\hat{\beta}_t + \hat{\sigma}_t a_i$.

The main implementation issue is the choice of $f_c(\cdot)$. We found it best to make the transformation $\log\hat{\sigma}$, giving the target density support $\mathbb{R}^{p+1}$, and to sample from a multivariate Student $t$ distribution with low degrees of freedom and with shape close to that of the target density. Details can be found in Brazzale (2000), Chapter 7.

The following two subsections summarize the results of a study inspired by Example 3 of DiCiccio, Field and Fraser (1990). The `rsm.sample` routine was used to retrieve the empirical distribution of the pivots $Z_1 = (\hat{\beta}-\beta)/\hat{\sigma}$ and $Z_2 = \hat{\sigma}/\sigma$ for a fixed value of the ancillary statistic, and to investigate the empirical accuracy of the tail area approximations (2) and (3).

## 5.3 Conditional Distribution of Pivotal Quantities

We considered a sample of size $n = 10$ from a linear regression model with the $10 \times 6$ design matrix

$$X^{\mathsf{T}} = \begin{bmatrix} 0.686 & 0.640 & 0.908 & 0.886 & 0.508 & 0.255 & 0.197 & 0.056 & 0.646 & 0.317 \\ 0.566 & 0.632 & 0.130 & 0.480 & 0.669 & 0.930 & 0.869 & 0.204 & 0.961 & 0.321 \\ 0 & 0 & 0 & 0 & 0 & 1 & 1 & 1 & 1 & 1 \\ 0 & 0 & 0 & 0 & 0 & 0.255 & 0.197 & 0.056 & 0.646 & 0.317 \\ 0 & 0 & 0 & 0 & 0 & 0.930 & 0.869 & 0.204 & 0.961 & 0.321 \\ 1 & 1 & 1 & 1 & 1 & 1 & 1 & 1 & 1 & 1 \end{bmatrix}$$

and errors that follow the standard log-Weibull distribution. This is a rather extreme scenario, with only 4 residual degrees of freedom and with highly correlated estimators of the regression coefficients. The sample configuration $a$ on which to condition was chosen by random sampling from the standard log-Weibull distribution using the same parameter values as in DiCiccio, Field and Fraser (1990). We repeated the study for various choices of $a$, all of which yielded similar results. The candidate generation density—a multivariate $t_5$ distribution—was rescaled and centered so as to optimize the acceptance rate, of about 25% and 30%, as was assessed in a pilot study. According to Corollary 4 of Tierney (1994), the resulting Markov chain is uniformly ergodic. The sampler was run for $T = 100{,}000$ iterations, reached stationarity very quickly, and mixed well. The corresponding `R` code is given in the demonstration file for the `csampling` package.

Figure 4 shows the conditional distributions of the pivots $Z_{13} = (\hat{\beta}_3 - \beta_3)/\hat{\sigma}$ and $\log Z_2 = \log(\hat{\sigma}/\sigma)$ for a particular choice of the sample configuration $a$. Both distributions are notably nonnormal; the finite sample distribution of $\log Z_2$ is, furthermore, heavily biased. Non-normal distributions were also observed for the remaining five regression coefficients. Table 3 compares the exact distribution functions of the two pivots with the approximations obtained from the likelihood root $r$ and the modified likelihood root $r^*$. The first order approximation performs rather poorly, especially for the scale parameter, while the third order solution competes well in this rather extreme scenario.

TABLE 3

*Exact and approximate tail probabilities for the pivots $Z_{13}$ and $\log Z_2$ in the log-Weibull regression model with $n = 10$ and $p = 6$ considered by DiCiccio, Field and Fraser (1990), Example 3, for a particular choice of $a$. The approximations considered are $\Phi(r)$ and $\Phi(r^*)$, where $r$ and $r^*$ are the likelihood root and its third order modification. The exact distribution was generated by the Metropolis–Hastings sampling (100,000 iterations, burn-in 5000)*

| | $\Pr(Z_{13} \le z)$ | | | | | | | |
|---|---|---|---|---|---|---|---|---|
| $z$ | $-52.3$ | $-38.6$ | $-29.0$ | $-20.1$ | $21.8$ | $30.6$ | $40.6$ | $55.3$ |
| exact | 0.01 | 0.025 | 0.05 | 0.1 | 0.9 | 0.95 | 0.975 | 0.99 |
| $\Phi(r)$ | $< 0.001$ | $< 0.001$ | 0.001 | 0.008 | 0.996 | 0.999 | 1.000 | 1.000 |
| $\Phi(r^*)$ | 0.006 | 0.016 | 0.035 | 0.082 | 0.913 | 0.962 | 0.984 | 0.995 |
| | $\Pr(\log Z_2 \le z)$ | | | | | | | |
| $z$ | $-2.13$ | $-1.89$ | $-1.69$ | $-1.46$ | $-0.30$ | $-0.16$ | $-0.05$ | $0.07$ |
| exact | 0.01 | 0.025 | 0.05 | 0.1 | 0.9 | 0.95 | 0.975 | 0.99 |
| $\Phi(r)$ | $< 0.001$ | $< 0.001$ | $< 0.001$ | $< 0.001$ | 0.170 | 0.295 | 0.433 | 0.586 |
| $\Phi(r^*)$ | 0.003 | 0.009 | 0.021 | 0.052 | 0.837 | 0.912 | 0.954 | 0.979 |

### 5.4 Accuracy of Higher Order Approximations

Table 4 of DiCiccio, Field and Fraser (1990) reports the overall rates of noncoverage of the one-sided confidence intervals obtained from $r$ and $\Phi^*$ for the parameters $\beta_1$, $\beta_3$ and $\log \sigma$ with a simulation of size 1000. However, as the authors themselves remark, these assessments are in terms of unconditional rather than conditional coverage.

Our Table 4 extends Table 4 of DiCiccio, Field and Fraser (1990): it gives the conditional rates of noncoverage of upper and lower confidence limits for the parameters $\beta_1$, $\beta_3$ and $\log \sigma$ obtained from the signed likelihood root pivot $r$, and the third order tail area approximations (2) and (3), for a particular choice of $a$. For the regression coefficients, the likelihood root yields confidence intervals which are too short, while the two higher order pivots work well. First order confidence intervals for $\log \sigma$ are heavily biased. Furthermore, we observed the feature mentioned by DiCiccio, Field and Fraser (1990): the tail area approximation $\Phi^*(r)$ breaks down. In about two-thirds of the samples the tail area exceeds 1. This is a drawback of Lugannani–Rice-type approximations, which need not give values within the interval $(0, 1)$. The modified likelihood root $r^*$ does not suffer from this drawback and provides satisfactory values. Some insight into why this happens is provided by Figure 5, which shows the normal Q–Q plots of $r$ and $r^*$ for $\beta_4$ and $\sigma$. The finite-sample distribution of $r(\sigma)$ is heavily biased with respect to the standard normal, whereas the tails of the distribution of $r(\beta_4)$ are too heavy. As noted in the previous paragraph, the conditional distributions of the maximum likelihood estimators are far from normal, so first order asymptotics are not useful. Surprisingly, $r^*$ works well for all seven parameters, especially since there are just $n = 10$ observations.

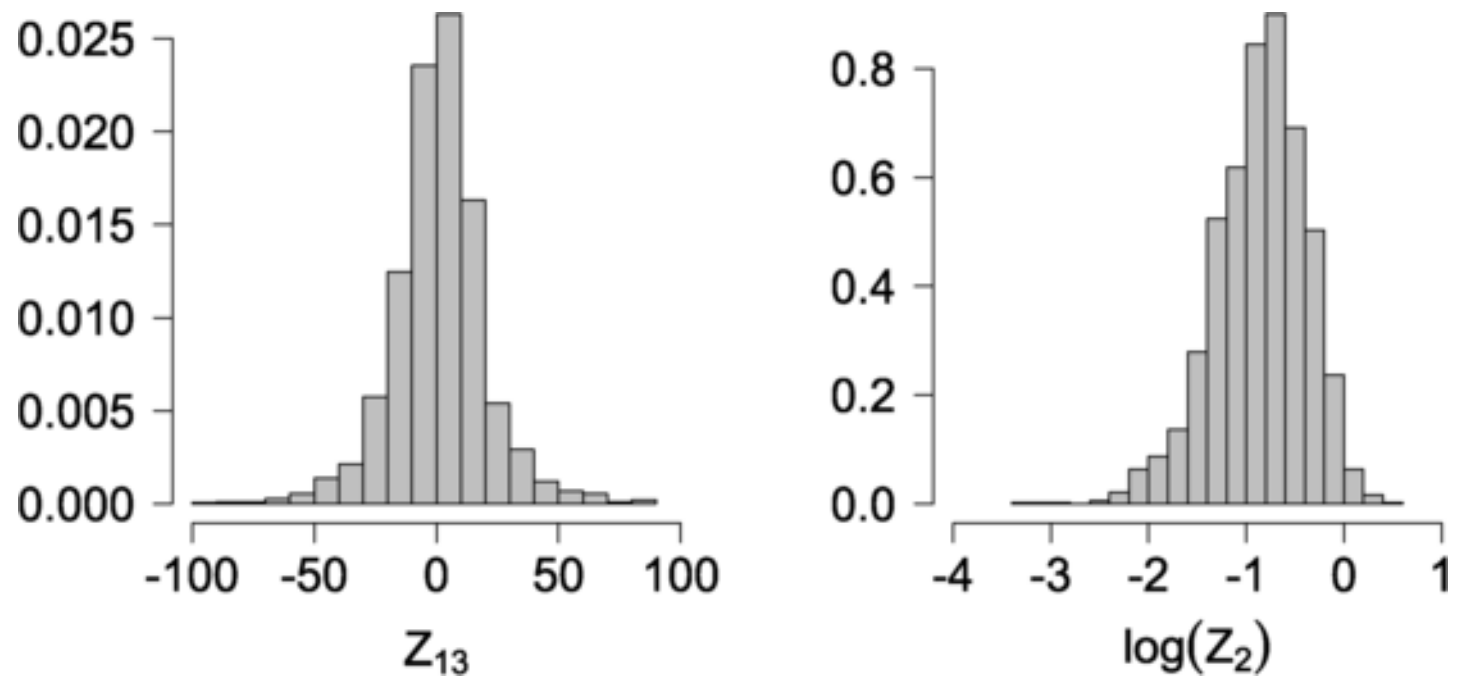

FIG. 4. *Histograms of the pivots $Z_{13}$ and $Z_2$ generated by the Metropolis–Hastings sampling (100,000 iterations). Only every 50th value is taken, after having discarded an initial sequence of length* 5000.

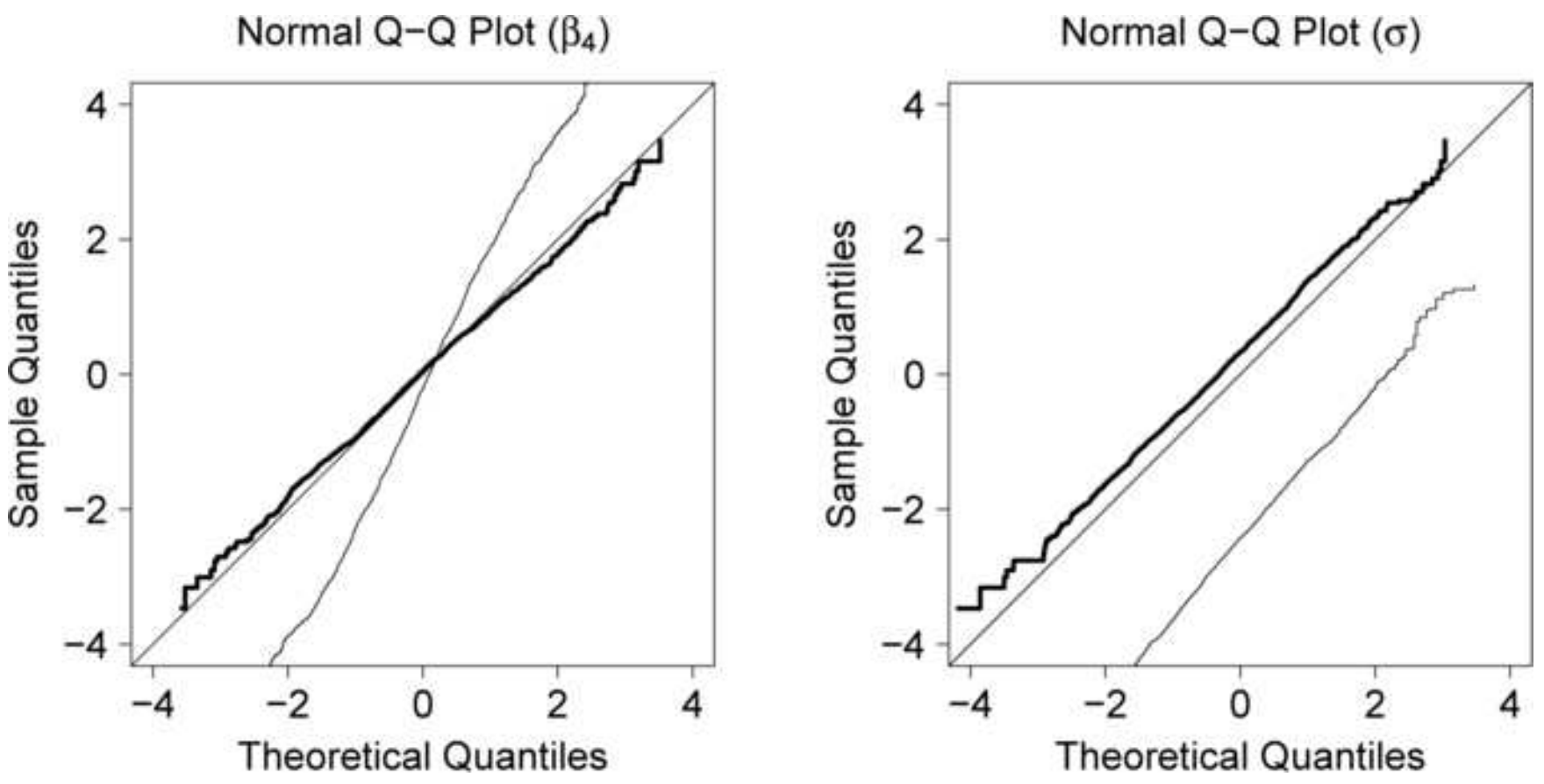

FIG. 5. *Normal Q–Q plots of $r$ (solid line) and $r^*$ (bold line) for $\beta_4$ (left panel) and $\sigma$ (right panel) obtained from the Metropolis–Hastings sampling (100,000 iterations), with diagonal line indicating perfect fit. The Q–Q plots only use every 50th simulated value, after having discarded an initial sequence of length* 5000*.*

## 6. NONLINEAR MODELS

Nonlinear models are widely used in applied statistics, especially for modeling dose-response curves. We consider the general form

$$y_{ij} = \mu(x_i; \beta) + w(x_i; \beta, \rho)\varepsilon_{ij}, \quad i = 1, \ldots, m, \; j = 1, \ldots, n_i, \tag{12}$$

where $m$ is the number of design points, $n_i$ the number of replicates at design point $x_i$, $y_{ij}$ represents the response of the $j$th experimental unit in the $i$th

TABLE 4

*Conditional rates of noncoverage of confidence intervals for $\beta_1$, $\beta_3$ and $\log \sigma$ in the log-Weibull regression model with $n = 10$ and $p = 6$ considered by DiCiccio, Field and Fraser (1990), Example 3, for a particular choice of $a$. The tail area approximations used are based on the likelihood root, $\Phi(r)$, on its higher order modification, $\Phi(r^*)$, and the third order quantity, $\Phi^*(r)$. The coverages were calculated using a Metropolis–Hastings chain of length* 100,000 *after having discarded the first* 5000 *values. The coverages of confidence intervals for $\log \sigma$ using $\Phi^*(r)$ are not given, as this statistic breaks down*

| | | Upper confidence limit | | | Lower confidence limit | | |
|---|---|---|---|---|---|---|---|
| | Nominal | $\Phi(r)$ | $\Phi(r^*)$ | $\Phi^*(r)$ | $\Phi(r)$ | $\Phi(r^*)$ | $\Phi^*(r)$ |
| $\beta_1$ | 0.5 | 7.96 | 0.91 | 0.41 | 11.16 | 0.89 | 0.33 |
| | 1 | 10.53 | 1.71 | 0.85 | 14.05 | 1.58 | 0.52 |
| | 2.5 | 15.28 | 3.65 | 2.24 | 18.79 | 3.36 | 1.29 |
| | 5 | 19.61 | 6.37 | 4.45 | 23.07 | 6.19 | 2.81 |
| | 10 | 25.19 | 11.88 | 9.77 | 28.87 | 11.76 | 6.60 |
| | 25 | 36.19 | 25.98 | 25.22 | 39.05 | 26.71 | 23.63 |
| $\beta_3$ | 0.5 | 8.29 | 0.86 | 0.35 | 10.89 | 1.03 | 0.40 |
| | 1 | 10.64 | 1.73 | 0.79 | 14.29 | 1.72 | 0.67 |
| | 2.5 | 15.15 | 3.89 | 2.22 | 19.23 | 3.69 | 1.69 |
| | 5 | 19.43 | 6.66 | 4.69 | 23.58 | 6.11 | 3.32 |
| | 10 | 25.24 | 11.82 | 9.63 | 29.61 | 11.30 | 6.76 |
| | 25 | 35.68 | 26.24 | 25.66 | 40.05 | 25.50 | 21.83 |
| $\log \sigma$ | 0.5 | 44.38 | 1.53 | – | 0.00 | 0.34 | – |
| | 1 | 53.46 | 2.69 | – | 0.00 | 0.51 | – |
| | 2.5 | 65.68 | 5.66 | – | 0.00 | 1.17 | – |
| | 5 | 75.14 | 9.53 | – | 0.01 | 2.64 | – |
| | 10 | 83.66 | 17.10 | – | 0.06 | 5.97 | – |
| | 25 | 93.46 | 35.88 | – | 0.40 | 16.85 | – |

group, and the errors $\varepsilon_{ij}$ are independent $N(0,1)$ variates. The mean response is given by the nonlinear function $\mu(x_i;\beta)$, which depends on a vector of unknown regression coefficients $\beta$, while the function $w(x_i;\beta,\rho)$ may also depend on a vector $\rho$ of variance parameters. If $w(\cdot)^2$ is constant, (12) becomes the classical nonlinear regression model. Inference on $\beta$ and $\rho$ is commonly based on first order approximations and linearization techniques (Seber and Wild (1989), Chapter 5), plus graphical summaries such as profile and contour plots (Bates and Watts (1988), Section 6.1), which allow one to assess the quality of distributional approximations for the likelihood ratio and Wald statistics. Bellio and Brazzale (1999) show that nonlinearity of the mean function and variance heterogeneity can lead to substantial inaccuracies in first order inferences, especially for the variance parameters, unless the sample size is large. This may be overcome using the higher order solutions presented in Section 2, which are relatively easily derived in this case (Bellio, Jensen and Seiden (2000)). To do so, we re-write (12) as a curved exponential family of order $(2m,d)$, where $d$ is the dimension of the parameter $\theta=(\beta,\rho)$. Expressions for the quantities needed to calculate the correction terms $q(\psi)$, $M(\psi)$ and $u(\psi)$ are listed in Brazzale, Davison and Reid (2007), Sections 8.6.2 and 8.6.3. We now present results of a data analysis performed with the `nlreg` package of the `hoa` bundle. Further examples may be found in Brazzale, Davison and Reid (2007), Chapters 5 and 6.

EXAMPLE 2 (*Herbicide bioassay*). Data set `C3` of the `nlreg` package concerns an in vitro bioassay on the action of the herbicide chlorsulfuron on the callus area of colonies of *Brassica napus L.*, also known as oilseed rape. The experiment is described in Seiden, Kappel and Streibig (1998) and consists of $n=51$ measurements of the callus `area` (in mm$^2$) for $m=10$ different `dose` levels (in nmol/l). The design is unbalanced, as the number of replicates per dose varies from 5 to 8. Bellio, Jensen and Seiden (2000) discuss a model where the response variable is the logarithm of the callus area and the mean function is the logarithm of the four-parameter logistic function

$$\mu(x;\beta)=\beta_1+\frac{\beta_2-\beta_1}{1+(x/\beta_4)^{\beta_3}},\qquad x\ge 0. \tag{13}$$

This yields a sigmoidal curve which decreases from an initial value $\beta_2$ to a limiting value $\beta_1$ when the concentration $x$ tends to infinity. The parameter $\beta_3$ determines the shape of the curve, and $\beta_4$ corresponds to the $EC_{50}$. A preliminary data analysis further suggests that the response variance might slightly decrease with its mean. The same authors suggest using the error-in-variable variance function

$$\begin{aligned} &w(x;\beta,\kappa,\gamma,\sigma^2)^2\\ &\quad=\sigma^2\left[1+\kappa x^{\gamma}\left\{\frac{\partial\mu(x;\beta)/\partial x}{\mu(x;\beta)}\right\}^2\right],\\ &\qquad\qquad\kappa,\gamma,\sigma^2>0,\end{aligned} \tag{14}$$

where $\kappa$, $\gamma$ and $\sigma^2$ are variance parameters.

The `R` code for first and higher order inference for the nonlinear model defined by (13) and (14) is

```
> C3.nl <-
+ nlreg(formula = log(area)~
+                 log(b1+(b2-b1)/
+                (1+(dose/b4)^b3))
+        weights = ~(1+((k*dose^g*
+                    (b2-b1)^2)/
+                (1+(dose/b4)^b3)^4*
+                 b3^2*dose^(2*b3-2))/
+                 b4^(2*b3)/(b1+(b2-b1)/
+                (1+(dose/b4)^b3))^2),
+        start = c(b1=2.2, b2=1700,
+                  b3=2.8, b4=0.28,
+                   g=2.7, k=1),
+        control = list(x.tol=1e-12,
+                       rel.tol=1e-12,
+                        step.min=1e-12),
+        data = C3, hoa = TRUE )
> summary( C3.nl )
> C3.prof <- profile( C3.nl,
+                       offset = "all" )
> contour( C3.prof, offset1 = b2,
+          offset2 = k, alpha = 0.95 )
> summary( ria.prof, twoside = TRUE )
```

The `formula` and `weights` arguments in the `nlreg` fitting routine determine respectively the mean and variance functions $\mu(\cdot)$ and $w(\cdot)^2$. The maximum likelihood estimate of $\sigma^2$ is available in closed form, but starting values for the other parameters must be provided through the `start` argument. All computations for $\sigma^2$ use the logarithmic scale. Because of the highly nonlinear model structure, we must refine the convergence criteria through the `control` argument. We obtain $\hat\beta_1=2.206$ (0.415), $\hat\beta_2=1662$ (117), $\hat\beta_3=2.841$ (0.360), $\hat\beta_4=0.2752$ (0.0452), $\hat\gamma=2.605$ (0.793), $\hat\kappa=1.009$ (0.580) and $\log\hat\sigma^2=-1.888$ (0.234). The values in brackets are the standard errors, as returned by a call to `summary`. The option

`hoa=TRUE` indicates that higher order solutions will be used in the subsequent calculations.

The `profile` and `contour` methods extend the original algorithm of Bates and Watts (1988), Chapter 6, to the higher order solutions presented in Section 2. The core routine is based upon pivot profiling as described in Section 3.3. By default, it computes the higher order statistics developed by Skovgaard (1996, 2001), although Fraser, Reid and Wu's (1999) version is available by setting `stats="fr"`. The option `offset="all"` means that all model parameters are to be profiled. Figure 6 shows the 95% approximate bivariate contour plots of the Wald, likelihood ratio and $w^*$ pivots for the parameters $\beta_2$ and $\kappa$. The contours are plotted on the original scale (right panel) and on the $r$ scale (left panel), respectively. On the latter scale the units are those of the likelihood root statistics. The more elliptical the contours are, the more quadratic is the likelihood; the closer the curves for $w$ and $w^*$, the better the behavior of first order inferences. The profile traces also shown represent the constrained maximum likelihood estimates of one parameter as a function of the other and show how the estimates affect each other. If the asymptotic correlation is zero, the angle between the traces is close to $\pi/2$, while an angle close to zero indicates strong correlation.

Two-sided confidence intervals can be obtained using the `summary` function. The 95% confidence intervals for the parameter $\kappa$ are $(-0.1270, 2, 145)$, $(0.2697, 3.096)$, $(0.3264, 3.546)$ and $(0.3321, 3.803)$ for respectively the Wald, $r$ and $r^*$ pivots, these last computed using Skovgaard's (1996) and Fraser, Reid and Wu (1999) formulation. Both versions of $r^*$ and the likelihood root, but not the Wald statistic, let us reject the null hypothesis of a constant variance function at the 5% level.

A common problem in nonlinear heteroscedastic regression is the estimation of the variance parameters, whose maximum likelihood estimators are usually heavily biased. More accurate estimates can be obtained from the adjusted profile likelihood using the fitting routine `mpl`:

```
> C3.mpl <- mpl( C3.nl )
> summary( C3.mpl )
```

We obtain $\hat{\gamma}_a = 2.654$ $(0.834)$, $\hat{\kappa}_a = 1.081$ $(0.711)$ and $\log\hat{\sigma}_a = -1.825$ $(0.248)$, when the correction term $M(\psi)$ is based upon the $p^*$ density approximation. The $p_{\mathrm{TEM}}$ approximation yields $\hat{\gamma}_a = 2.650$ $(0.833)$, $\hat{\kappa}_a = 1.092$ $(0.721)$ and $\log\hat{\sigma}_a = -1.827$ $(0.248)$. The standard errors are obtained from the profile information matrix $j_{\mathrm{p}}(\hat{\psi})$; this is possible because $|j_{\mathrm{a}}(\hat{\psi})| = |j_{\mathrm{p}}(\hat{\psi})|\{1+O_p(n^{-1})\}$ and $\hat{\psi}_{\mathrm{a}} - \hat{\psi} = O_p(n^{-1})$. The distance between the values obtained from the profile likelihood and the adjusted profile likelihood functions gives an idea of the bias of the ordinary maximum likelihood estimators. It is reassuring that both versions of the adjusted profile likelihood yield similar estimates.

## 7. DISCUSSION

The purpose of this paper is to show that highly accurate likelihood methods may be routinely used in data analysis, both to check whether standard approximations are adequate and to supplement them when they are inadequate. Libraries are available that implement these procedures for a variety of common models. Although available for the numerical computing environment R, it would be relatively straightforward to modify them for other packages. The classes of models discussed in this paper form a small subset of those used in practice, but the same ideas can be extended to other classes for which some unifying structure can be identified, so that a common statistical computation setup is possible; see, for example, Guolo, Brazzale and Salvan (2006). In other cases a general approach requiring a small amount of programming is described by Brazzale, Davison and Reid (2007), Section 9.5.

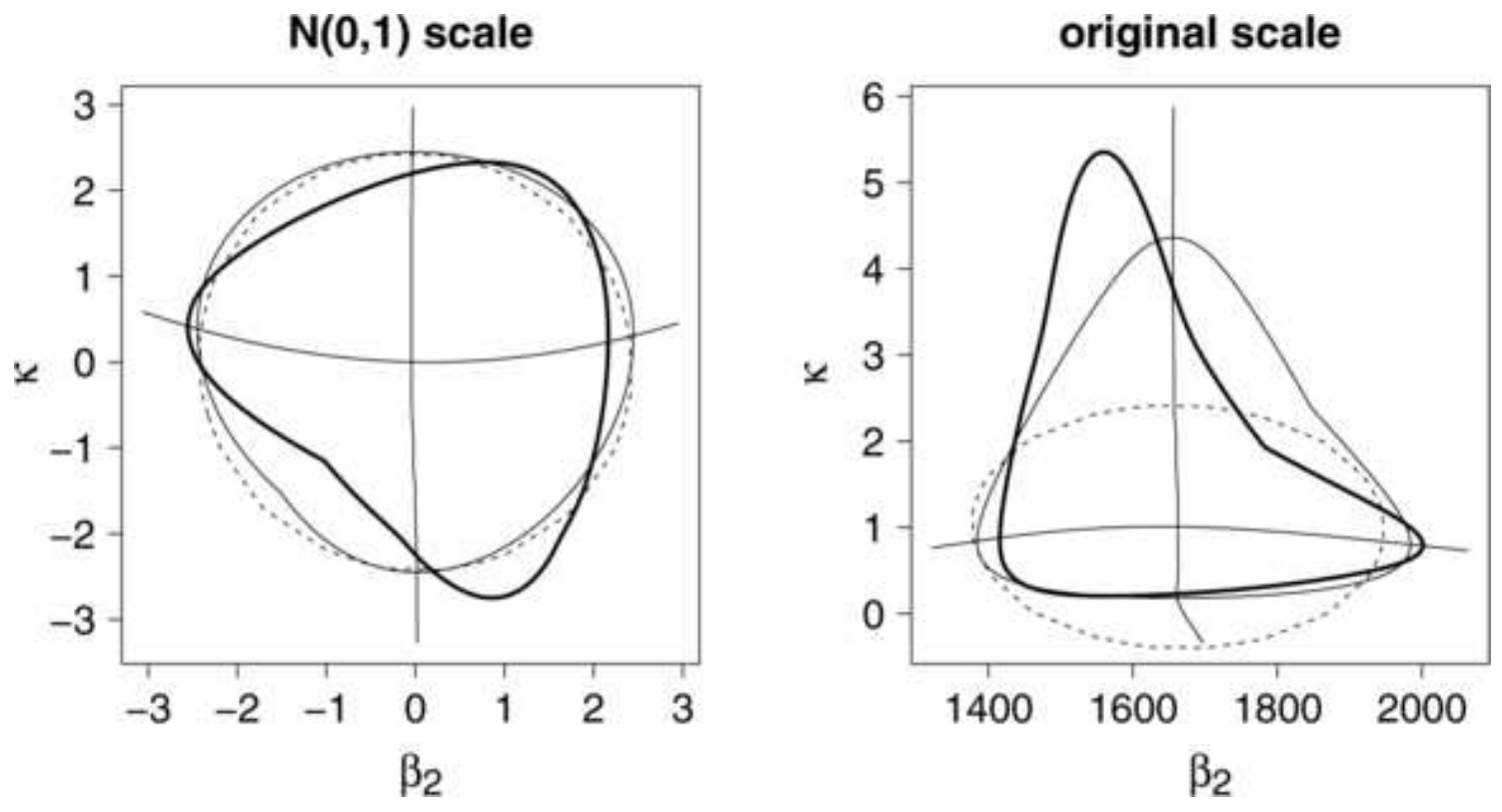

FIG. 6. *Herbicide bioassay: data set* `C3`. *Approximate bivariate contour plots and profile traces for the parameters $\beta_2$ and $\kappa$ obtained with the* `contour` *method of the* `nlreg` *package ($\alpha = 0.05$). The pivots used are as follows: likelihood root (—), modified likelihood root $w^*$ (—) and Wald (- - -). Right panel: original scale; left panel: $r$ scale.*

In our work we have focused on so-called double saddlepoint approximations, although sequential saddlepoint approximations leading to inference based on expressions such as (5) should also be mentioned (Pierce and Peters (1992); DiCiccio and Martin (1993)). The maximum likelihood estimator $\hat{\psi}_{\mathrm{a}}$ provides good point estimates, for example, when estimating the variance parameters in a linear or nonlinear regression model; the bias correction is essentially what is provided by the use of REML through maximization of the restricted likelihood function. If an adjusted profile log likelihood derived from (5) is available, comparison of it with the corresponding profile log likelihood $\ell_{\mathrm{p}}(\psi)$ gives valuable information about the bias of the maximum likelihood estimator, and if the adjusted profile log likelihood, $\log L_{\mathrm{a}}(\psi)$, is close to quadratic, then it should be safe to base inference for $\psi$ on the corresponding likelihood root statistic $r_{\mathrm{a}}(\psi)$. However, if the profile log likelihoods are asymmetric, then inference based on $r^*(\psi)$ or, if available, $r^*_{\mathrm{a}}(\psi)$, will be preferable. For instance, the asymmetry of both curves in the left-hand panel of Figure 1 is a warning to avoid using Wald statistics.

Different expressions are available for the correction terms $q(\psi)$ and $M(\psi)$ in (4) and (5). Though almost equivalent from the analytical and the numerical points of view, preference for one version or the other is not merely a matter of taste. Expression (15) in Appendix A.1 requires the availability of an exact ancillary statistic, and this is rarely the case. Skovgaard's (1996) sample space approximations circumvent this problem, but require the calculation of the covariances that are involved. Expression (16) of the Appendix is more versatile both in its derivation and implementation, but it seems unclear if it applies to dependent data. Bellio and Sartori (2006) illustrate the versatility of the higher order methods discussed in this paper.

A parallel literature on analytical approximations for Bayesian inference based on marginal posterior densities leads to remarkably similar expressions. Both the conceptual and mathematical developments are simpler, the first because arguments involving ancillarity are not required in the Bayesian paradigm, and the second because Laplace approximation for integrals is used (Tierney and Kadane (1986), Tierney, Kass and Kadane (1989), DiCiccio, Field and Fraser (1990)). Elementary expositions are given by Davison (2003), Section 11.3.1 and Brazzale, Davison and Reid (2007), Section 8.7. The relation with matching and noninformative priors (Tibshirani (1989), Reid, Mukerjee and Fraser (2002)) yields a close but imperfect rapprochement between objective Bayesian and Fisherian approaches. If so, the Holy Grail of objective statistical inference sought by Jeffreys and Fisher will have been reached—at least approximately! A recent example of this rapprochement is given by Davison and Sartori (2008).

Almost all the literature on higher order likelihood inference concerns regular models, yet nonregular situations are of growing interest. Testing for zero variance components plays a role in both spline smoothing applications and in generalized linear mixed models, for example, and the boundary hypotheses this entails lead to modifications of the usual limiting distributions. It would be valuable to have accurate and practicable analytical approximations for the more common nonregular situations. A first step in this direction is made by Castillo and López-Ratera (2006).

We have mainly discussed analytical approximations, but DiCiccio, Martin and Stern (2001), Lee and Young (2005), and DiCiccio and Young (2008) have used the parametric bootstrap to achieve high accuracy. As yet, the properties of this approach are understood only in certain, albeit important, cases, and rather large simulations seem to be needed for it to give solid gains over analytical approximation, which we believe to be sufficiently accurate for most applications. Differences in the first two decimal places of a P-value may influence decisions taken in practice, while variation in further places is crucial only in exceptional cases.

In this paper our concern is with implementation of accurate likelihood inference, which typically entails the elimination of parameters from likelihoods, by appropriate, often approximate, conditioning or marginalization. The different roles of conditioning in inference have been aired at length in the literature, and the interested reader may refer to Cox (1988) or Reid (1995), for instance, for more general discussion on this topic. The discussion of Section 4 might be misinterpreted as an attack on conditional inference, but it is rather intended to point out that the properties of statistical procedures labelled 'exact' merit critical examination. So-called exact inference may come at too high a price.

Why seek highly accurate inferences for a model that may be incorrect? Although we entirely agree that the robustness of conclusions plays a key role in applied work, we believe that this question is slightly beside the point. Provided the assumed model is found empirically acceptable, checking and, if necessary, improving the usual basis for inference seems worthwhile, even if the model might be falsified based on a larger sample. A Jesuitical reply might be: faced with an apparently Gaussian sample of size ten, should one base inference for its mean on a normal approximation or on the Student $t$ distribution? A reader who would opt for the latter should also be willing on occasion to use the approaches described above. It would be worrying if the higher order methods were very sensitive to model failure, but published and unpublished work on this (Fraser, Wong and Wu (1999), Bellio (2000)), as well as our own limited simulations, suggest that there is no dramatic breakdown of them under small model perturbations. It would be reassuring to have more evidence of this, however.

We hope that this paper will encourage others to use higher order methods in their applied work: a recipe may be appetizing in theory, but the proof of a pudding is in the eating.

## APPENDIX: HIGHER ORDER APPROXIMATIONS

### A.1 Modified Likelihood Root

The key element in using the modified likelihood root is the form of (4), whether computed using $\varphi(\theta)$ or using equivalent expressions.

Barndorff-Nielsen (1983) gives

$$q_1 = \frac{|\ell_{;\hat\theta}(\hat\theta) - \ell_{;\hat\theta}(\hat\theta_\psi)\ \ell_{\lambda;\hat\theta}(\hat\theta_\psi)|}{|\ell_{\theta;\hat\theta}(\hat\theta)|} \cdot \left\{\frac{|j(\hat\theta)|}{|j_{\lambda\lambda}(\hat\theta_\psi)|}\right\}^{1/2}, \tag{15}$$

where the data vector $y$ in $\ell(\theta;y)$ is expressed as a one-to-one function of the maximum likelihood estimator $\hat\theta$ and of an ancillary statistic $a$, whose distribution does not depend on $\theta$. Expression (15) is obtained by formally setting $\varphi(\theta)^\top = \ell_{;\hat\theta}(\theta;\hat\theta,a)$. The numerator on the right-hand side of (15) is the determinant of a $d\times d$ matrix whose first column is the difference of sample-space derivatives $\ell_{;\hat\theta}(\hat\theta) - \ell_{;\hat\theta}(\hat\theta_\psi)$, defined as $\ell_{;\hat\theta}(\theta) = \ell_{;\hat\theta}(\theta;\hat\theta,a) = \partial\ell(\theta;\hat\theta,a)/\partial\hat\theta$, and whose remaining columns comprise the $d\times(d-1)$ matrix of mixed derivatives $\ell_{\lambda;\hat\theta}(\theta) = \partial^2\ell(\psi,\lambda;\hat\theta,a)/\partial\hat\theta\,\partial\lambda^\top$ evaluated at the constrained estimate $\hat\theta_\psi$. The denominator contains the $d\times d$ matrix of mixed derivatives $\ell_{\theta;\hat\theta}(\theta) = \partial^2\ell(\theta;\hat\theta,a)/\partial\hat\theta\,\partial\theta^\top$ evaluated at $\hat\theta$.

A difficulty in using (15) is the need to differentiate $\ell(\theta;y)$ partially with respect to the maximum likelihood estimator $\hat\theta$, while holding fixed the value of a full-dimensional ancillary statistic $a$. Fraser, Reid and Wu (1999) bypass this difficulty in several ways. First, they note that only a second order, approximate, ancillary is needed, so the results apply to a broader range of models. Second, the ancillary is needed only at the observed data point, and this is given by tangents $(V_1,\dots,V_d)$ defined in the directions corresponding to fixed values of the ancillary. Third, any differentiation in these directions is allowed, but differentiation with respect to $\hat\theta$ is irrelevant. Thus, the nominal reparameterization can be given as $\varphi(\theta)^\top = \ell_{;V}(\theta;y)$ using directional derivatives and yielding an expression which compares directly with (15),

$$q_2 = \frac{|\ell_{;V}(\hat\theta) - \ell_{;V}(\hat\theta_\psi)\ \ell_{\lambda;V}(\hat\theta_\psi)|}{|\ell_{\theta;V}(\hat\theta)|} \cdot \left\{\frac{|j(\hat\theta)|}{|j_{\lambda\lambda}(\hat\theta_\psi)|}\right\}^{1/2}. \tag{16}$$

Here, $\varphi(\theta)$ is the canonical parameter of an exponential family with sufficient statistic $s = s(y) = \partial\ell(\hat\theta^0;y)/\partial\theta$, the score function evaluated at the maximum likelihood estimate $\hat\theta^0$ for a fixed point $y^0$, which approximates the true model locally at $y^0$ with a relative error of the order $O(n^{-3/2})$ and whose log likelihood function and first derivative with respect to $\theta$ at the fixed point $y^0$ equal those of the original model. The canonical parameter $\varphi$ is defined using a set of $n$ vectors of length $d$ through an $n\times d$ matrix $V$, with rows $V_1, \dots, V_n$, where $\ell_{;V}(\theta; y)$ indicates that the log likelihood is differentiated on the surface spanned by the columns of $V$. If the observations $y_1, \dots, y_n$ are independent, then $\ell_{;V}(\theta; y) = \sum_{i=1}^n V_i \partial\ell(\theta; y)/\partial y_i$. In the continuous case, the $V_i$ can be constructed as

$$V = -\left(\frac{\partial z}{\partial y^\top}\right)^{-1}\left(\frac{\partial z}{\partial \theta^\top}\right)\bigg|_{\theta=\hat\theta},$$

using a vector of pivotal quantities $z = \{z_1(y_1, \theta), \dots, z_n(y_n, \theta)\}$, where each component $z_i(y_i, \theta)$ has a fixed distribution under the model. In the continuous case such a vector always exists in the form of the probability integral transformation $F(y_i; \theta)$.

Skovgaard (1996) develops an approximation to $q_1$, which avoids specification of $a$ by approximating the sample space derivatives in (15) as

$$\ell_{;\hat\theta}(\hat\theta) - \ell_{;\hat\theta}(\hat\theta_\psi) \doteq \{i(\hat\theta)\}^{-1} j(\hat\theta) Q(\hat\theta, \hat\theta_\psi),$$
$$\ell_{\theta;\hat\theta}(\hat\theta_\psi) \doteq \{i(\hat\theta)\}^{-1} j(\hat\theta) S(\hat\theta, \hat\theta_\psi),$$

using moments of quantities such as the expected Fisher information $i(\theta)$ and the covariances

$$S(\theta_1, \theta_2) = \mathrm{cov}_{\theta_1}\{\ell_\theta(\theta_1), \ell_\theta(\theta_2)\},$$
$$Q(\theta_1, \theta_2) = \mathrm{cov}_{\theta_1}\{\ell_\theta(\theta_1), \ell(\theta_1) - \ell(\theta_2)\}.$$

The covariances $S$ and $Q$ are often readily computed, either analytically or by simulation, though the resulting tail area approximation has error $O(n^{-1})$ rather than $O(n^{-3/2})$. In an as-yet unpublished work, Fraser and Reid (2009) point out that this version may be obtained by replacing $\varphi(\theta)$ by $\partial E\{\ell(\theta; y); \theta_0\}/\partial\theta_0$, evaluated at $\theta_0 = \hat\theta$.

### A.2 Adjusted Profile Likelihood

The correction term $M(\psi)$ can be derived using the $p^*$ approximation,

$$M_1(\psi) = |j_{\lambda\lambda}(\hat\theta_\psi)|^{1/2}/|\ell_{\lambda;\hat\lambda}(\hat\theta_\psi)|,$$

which gives rise to Barndorff-Nielsen (1983) modified profile likelihood, or using the tangent exponential model,

$$M_2(\psi) = |j_{\lambda\lambda}(\hat\theta_\psi)|^{1/2}/|\varphi_\lambda(\hat\theta_\psi)^\top \hat V_\lambda|,$$

where $\hat V_\lambda$ is the $n \times (d - d_0)$ matrix obtained from $\hat V = V(\hat\theta)$ by omitting the columns which relate to the parameter of interest (Fraser (2003)).

## ACKNOWLEDGMENTS

We thank Ruggero Bellio, Donald Fraser, Jon Forster, Nancy Reid, Alessandra Salvan, Nicola Sartori and two referees for very helpful comments and suggestions. We acknowledge funding from the Swiss National Science Foundation in the context of the National Centre for Competence in Research in Plant Survival (www.unine.ch/nccr) and from the Ministero dell'Istruzione, dell'Università e della Ricerca, Italy. The first author benefitted from a postdoctoral fellowship at the Department of Statistics, University of Toronto.

## REFERENCES

AGRESTI, A. (2002). *Categorical Data Analysis*, 2nd ed. Wiley, New York. MR1914507

AGRESTI, A. and CAFFO, B. (2000). Simple and effective confidence intervals for proportions and differences of proportions result from adding two successes and two failures. *Amer. Statist.* **54** 280–288. MR1814845

AGRESTI, A. and COULL, B. A. (1998). Approximate is better than "exact" for interval estimation of binomial proportions. *Amer. Statist.* **52** 119–126. MR1628435

ANDREWS, D. F. and HERZBERG, A. M. (1985). *Data: A Collection of Problems from Many Fields for the Student and Research Worker*. Springer, New York.

BARNDORFF-NIELSEN, O. E. (1983). On a formula for the distribution of the maximum likelihood estimator. *Biometrika* **70** 343–365. MR0712023

BARNDORFF-NIELSEN, O. E. (1986). Inference on full or partial parameters based on the standardized signed log likelihood ratio. *Biometrika* **73** 307–322. MR0855891

BARNDORFF-NIELSEN, O. E. and COX, D. R. (1979). Edgeworth and saddle-point approximations with statistical applications (with discussion). *J. Roy. Statist. Soc. Ser. B* **41** 279–312. MR0557595

BARNDORFF-NIELSEN, O. E. and COX, D. R. (1994). *Inference and Asymptotics*. Chapman & Hall, London. MR1317097

BATES, D. M. and WATTS, D. G. (1988). *Nonlinear Regression Analysis and Its Applications*. Wiley, New York. MR1060528

BELLIO, R. (2000). Likelihood asymptotics: Applications in biostatistics. Ph.D. thesis, Dept. Statistics, Univ. Padova, Italy.

BELLIO, R. and BRAZZALE, A. R. (1999). Higher-order asymptotics in nonlinear regression. In *Proceedings of the 14th International Workshop on Statistical Modelling, Graz, July 19–23, 1999* (H. Friedl, A. Berghold and G. Kauermann, eds.) 440–443.

BELLIO, R. and BRAZZALE, A. R. (2003). Higher-order asymptotics unleashed: Software design for nonlinear heteroscedastic models. *J. Comput. Graph. Statist.* **12** 682–697. MR2005458

BELLIO, R., JENSEN, J. E. and SEIDEN, P. (2000). Applications of likelihood asymptotics for nonlinear regression in herbicide bioassays. *Biometrics* **56** 1204–1212. MR1815621

Bellio, R. and Sartori, N. (2006). Practical use of modified maximum likelihoods for stratified data. *Biom. J.* **48** 876–886. MR2291296

Bondesson, L. (1982). To reduce a composite hypothesis to a simple one by sampling from the structural or a conditional distribution. *Scand. J. Statist.* **9** 129–138. MR0680908

Brazzale, A. R. (1999). Approximate conditional inference in logistic and loglinear models. *J. Comput. Graph. Statist.* **8** 653–661.

Brazzale, A. R. (2000). Practical small-sample parametric inference. Ph.D. thesis, Dept. Mathematics, Swiss Federal Institute of Technology Lausanne. Available at www.isib.cnr.it/~brazzale/lib.html.

Brazzale, A. R. (2005). hoa: An R package bundle for higher order likelihood inference. *R News* **5** 20–27. Available at http://cran.r-project.org/doc/Rnews.

Brazzale, A. R., Davison, A. C. and Reid, N. (2007). *Applied Asymptotics: Case Studies in Small Sample Statistics*. Cambridge Univ. Press. MR2342742

Brown, L. D., Cai, T. T. and DasGupta, A. (2001). Interval estimation for a binomial proportion (with discussion). *Statist. Sci.* **16** 101–133. MR1861069

Butler, R. W. (2007). *Saddlepoint Approximations with Applications*. Cambridge Univ. Press. MR2357347

Casella, G., Wells, M. T. and Tanner, M. A. (1994). Using sampling-based calculations for pivotal inference. Technical Report BU-1178-M, Biometrics Unit, Cornell Univ., Ithaca, NY.

Castillo, J. D. and López-Ratera, A. (2006). Saddlepoint approximation in exponential models with boundary points. *Bernoulli* **12** 491–500. MR2232728

Clopper, C. J. and Pearson, E. S. (1934). The use of confidence interval or fiducial limits illustrated in the case of the binomial. *Biometrika* **26** 404–413.

Cox, D. R. (1958). The regression analysis of binary sequences (with discussion). *J. Roy. Statist. Soc. Ser. B* **20** 215–242. MR0099097

Cox, D. R. (1988). Some aspects of conditional and asymptotic inference: A review. *Sankhyà A* **50** 314–337. MR1065547

Cytel Inc. (2007). *StatXact/LogXact 8*. Cytel Inc., Cambridge, Mas. Available at http://www.cytel.com.

Daniels, H. E. (1954). Saddlepoint approximations in statistics. *Ann. Math. Statist.* **25** 631–650. MR0066602

Daniels, H. E. (1958). Discussion of "The regression analysis of binary sequences," by D. R. Cox. *J. Roy. Statist. Soc. Ser. B* **20** 236–238. MR0099097

Daniels, H. E. (1987). Tail probability approximations. *Internat. Statist. Rev.* **54** 37–48. MR0962940

Davison, A. C. (1988). Approximate conditional inference in generalized linear models. *J. Roy. Statist. Soc. Ser. B* **50** 445–461. MR0970979

Davison, A. C. (2003). *Statistical Models*. Cambridge Univ. Press. MR1998913

Davison, A. C., Fraser, D. A. S. and Reid, N. (2006). Improved likelihood inference for discrete data. *J. Roy. Statist. Soc. Ser. B* **68** 495–508. MR2278337

Davison, A. C. and Hinkley, D. V. (1997). *Bootstrap Methods and Their Application*. Cambridge Univ. Press. MR1478673

Davison, A. C. and Sartori, N. (2008). The Banff challenge: Statistical detection of a noisy signal. *Statist. Sci.* **123** 354–364.

Davison, A. C. and Wang, S. (2002). Saddlepoint approximations as smoothers. *Biometrika* **89** 933–938. MR1946521

Denison, D. G. T., Holmes, C. C., Mallick, B. K. and Smith, A. F. M. (2002). *Bayesian Methods for Nonlinear Classification and Regression*. Wiley, New York. MR1962778

Diaconis, P. and Sturmfels, B. (1998). Algebraic algorithms for sampling from conditional distributions. *Ann. Statist.* **26** 363–397. MR1608156

DiCiccio, T. J. and Efron, B. (1996). Bootstrap confidence intervals (with discussion). *Statist. Sci.* **11** 189–228. MR1436647

DiCiccio, T. J. and Field, C. A. (1991). An accurate method for approximate conditional and Bayesian inference about linear regression models from censored data. *Biometrika* **78** 903–910. MR1147027

DiCiccio, T. J., Field, C. A. and Fraser, D. A. S. (1990). Approximations of marginal tail probabilities and inference for scalar parameters. *Biometrika* **77** 77–95. MR1049410

DiCiccio, T. J. and Martin, M. A. (1993). Simple modifications for signed roots of likelihood ratio statistics. *J. Roy. Statist. Soc. Ser. B* **55** 305–316. MR1210437

DiCiccio, T. J., Martin, M. A. and Stern, S. E. (2001). Simple and accurate one-sided inference from signed roots of likelihood ratios. *Canad. J. Statist.* **29** 67–76. MR1834487

DiCiccio, T. J. and Young, G. A. (2008). Conditional properties of unconditional parametric bootstrap procedures for inference in exponential families. *Biometrika* **95** 747–758.

Durbin, J. (1961). Some methods of constructing exact tests. *Biometrika* **48** 41–55. MR0126313

Efron, B. (1979). Bootstrap methods: Another look at the jackknife. *Ann. Statist.* **7** 1–26. MR0515681

Efron, B. and Hinkley, D. V. (1978). Assessing the accuracy of the maximum likelihood estimator: Observed versus expected Fisher information (with discussion). *Biometrika* **65** 457–487. MR0521817

Efron, B. and Tibshirani, R. J. (1993). *An Introduction to the Bootstrap*. Chapman & Hall, New York. MR1270903

Fisher, R. A. (1934). Two new properties of mathematical likelihood. *Proc. R. Soc. Lond. Ser. A* **144** 285–307.

Forster, J. J., McDonald, J. W. and Smith, P. W. F. (1996). Monte Carlo exact conditional tests for log-linear and logistic models. *J. Roy. Statist. Soc. Ser. B* **58** 445–453. MR1377843

Forster, J. J., McDonald, J. W. and Smith, P. W. F. (2003). Markov chain Monte Carlo exact inference for binomial and multinomial regression models. *Stat. Comput.* **13** 169–177. MR1963333

Fraser, D. A. S. (1979). *Inference and Linear Models*. McGraw-Hill, New York. MR0535612

Fraser, D. A. S. (1990). Tail probabilities from observed likelihoods. *Biometrika* **77** 65–76. MR1049409

Fraser, D. A. S. (2003). Likelihood for component parameters. *Biometrika* **90** 327–339. MR1986650

Fraser, D. A. S., Andrews, D. A. and Wong, A. (2005). Computation of distribution functions from likelihood in-formation near observed data. *J. Statist. Plann. Inference* **134** 180–193. MR2146092

Fraser, D. A. S., Lee, H. S. and Reid, N. (1990). Nonnormal linear regression: An example of significance levels in high dimensions. *Biometrika* **77** 333–341. MR1064805

Fraser, D. A. S. and Reid, N. (2009). Mean likelihood and higher order approximations. Unpublished.

Fraser, D. A. S., Reid, N. and Wu, J. (1999). A simple formula for tail probabilities for frequentist and Bayesian inference. *Biometrika* **86** 249–264. MR1705367

Fraser, D. A. S., Wong, A. and Wu, J. (1999). Regression analysis, nonlinear or nonnormal: Simple and accurate $p$ values from likelihood analysis. *J. Amer. Statist. Assoc.* **94** 1286–1295. MR1731490

Geyer, C. J. and Meeden, G. D. (2005). Fuzzy and randomized confidence intervals and P-values (with discussion). *Statist. Sci.* **20** 358–387. MR2210225

Guolo, A., Brazzale, A. R. and Salvan, A. (2006). Improved inference on a scalar fixed effect of interest in nonlinear mixed-effects models. *Comput. Statist. Data Anal.* **51** 1602–1613. MR2307530

Jensen, J. L. (1992). The modified signed likelihood statistic and saddlepoint approximations. *Biometrika* **79** 693–703. MR1209471

Kappenmann, R. F. (1975). Conditional confidence intervals for the double exponential distribution parameters. *Technometrics* **17** 233–235. MR0370902

Lawless, J. F. (1972). Conditional confidence interval procedures for the location and scale parameters of the Cauchy and logistic distributions. *Biometrika* **59** 377–386. MR0334380

Lawless, J. F. (1973). Conditional versus unconditional confidence intervals for the parameters of the Weibull distribution. *J. Amer. Statist. Assoc.* **68** 655–669.

Lawless, J. F. (1978). Confidence interval estimation for the Weibull and extreme value distributions. *Technometrics* **20** 355–368. MR0515992

Lee, S. M. S. and Young, G. A. (2005). Parametric bootstrapping with nuisance parameters. *Statist. Probab. Lett.* **71** 143–153. MR2126770

Lugannani, R. and Rice, S. (1980). Saddle point approximation for the distribution of the sum of independent random variables. *Adv. Appl. Probab.* **12** 475–490. MR0569438

Lunn, D. J., Thomas, A., Best, N. and Spiegelhalter, D. (2000). WinBUGS—a Bayesian modelling framework: Concepts, structure, and extensibility. *Stat. Comput.* **10** 325–337. Available at `http://www.mrc-bsu.cam.ac.uk/bugs`.

Mehta, C. R. and Patel, N. R. (1995). Exact logistic regression: Theory and examples. *Stat. Med.* **14** 2143–2160.

Mehta, C. R., Patel, N. R. and Senchaudhuri, P. (2000). Efficient Monte Carlo methods for conditional logistic regression. *J. Amer. Statist. Assoc.* **95** 99–108.

Morgenthaler, S. and Tukey, J. W., eds. (1991). *Configural Polysampling: A Route to Practical Robustness*. Wiley, New York. MR1247031

Pace, L. and Salvan, A. (1997). *Principles of Statistical Inference from a Neo-Fisherian Perspective*. World Scientific, Singapore. MR1476674

Pace, L., Salvan, A. and Ventura, L. (2004). The effects of rounding on likelihood procedures. *J. Appl. Statist.* **31** 29–48. MR2041554

Pierce, D. A. and Peters, D. (1992). Practical use of higher order asymptotics for multiparameter exponential families (with discussion). *J. Roy. Statist. Soc. Ser. B* **54** 701–737. MR1185218

Pierce, D. A. and Peters, D. (1999). Improving on exact tests by approximate conditioning. *Biometrika* **86** 265–277. MR1705363

R Development Core Team (2007). *R: A Language and Environment for Statistical Computing*. R Foundation for Statistical Computing, Vienna, Austria. ISBN 3-900051-07-0. Available at `http://www.R-project.org`.

Reid, N. (1988). Saddlepoint methods and statistical inference (with discussion). *Statist. Sci.* **3** 213–238. MR0968390

Reid, N. (1995). The roles of conditioning in inference (with discussion). *Statist. Sci.* **10** 138–199. MR1368097

Reid, N. (2003). Asymptotics and the theory of inference. *Ann. Statist.* **31** 1695–1731. MR2036388

Reid, N., Mukerjee, R. and Fraser, D. A. S. (2002). Some aspects of matching priors. In *Mathematical Statistics and Applications: Festschrift for Constance van Eeden* (M. Moore, S. Froda and C. Léger, eds.) 31–44. Hayward, CA. MR2138284

Robert, C. P. and Casella, G. (2004). *Monte Carlo Statistical Methods*, 2nd ed. Springer, New York. MR2080278

Ronchetti, E. and Ventura, L. (2000). Between stability and higher-order asymptotics. *Stat. Comput.* **11** 67–73. MR1837146

S-PLUS (2007). *S-PLUS® (version 8)*. Insightful Corporation, Seattle, WA. Available at `http://www.insightful.com`.

Sartori, N. (2003). Modified profile likelihoods in models with stratum nuisance parameters. *Biometrika* **90** 533–549. MR2006833

Sartori, N., Bellio, R., Pace, L. and Salvan, A. (1999). The directed modified profile likelihood in models with many nuisance parameters. *Biometrika* **86** 735–742. MR1723792

Seber, G. A. F. and Wild, A. J. (1989). *Nonlinear Regression*. Wiley, New York. MR0986070

Seiden, P., Kappel, D. and Streibig, J. C. (1998). Response of *Brassica napus L.* tissue culture to metsulfuron methyl and chlorsulfuron. *Weed Research* **38** 221–228.

Severini, T. A. (1999). An empirical adjustment to the likelihood ratio statistic. *Biometrika* **86** 235–247. MR1705371

Severini, T. A. (2000). *Likelihood Methods in Statistics*. Clarendon, Oxford. MR1854870

Skovgaard, I. M. (1987). Saddlepoint expansions for conditional distributions. *J. Appl. Probab.* **24** 875–887. MR0913828

Skovgaard, I. M. (1996). An explicit large-deviation approximation to one-parameter tests. *Bernoulli* **2** 145–166. MR1410135

Skovgaard, I. M. (2001). Likelihood asymptotics. *Scand. J. Statist.* **28** 3–32. MR1844348

Smith, P. W. F., Forster, J. J. and McDonald, J. W. (1996). Monte Carlo exact tests for square contingency tables. *J. Roy. Statist. Soc. Ser. A* **159** 309–321.

Strawderman, R. L., Casella, G. and Wells, M. T. (1996). Practical small-sample asymptotics for regression problems (with discussion). *J. Amer. Statist. Assoc.* **91** 643–654. MR1395732

Tibshirani, R. J. (1989). Noninformative priors for one parameter of many. *Biometrika* **76** 604–608. MR1040654

Tierney, L. (1994). Markov chains for exploring posterior distributions (with discussion). *Ann. Statist.* **22** 1701–1762. MR1329166

Tierney, L. and Kadane, J. B. (1986). Accurate approximations for posterior moments and marginal densities. *J. Amer. Statist. Assoc.* **81** 82–86. MR0830567

Tierney, L., Kass, R. E. and Kadane, J. B. (1989). Approximate marginal densities of nonlinear functions. *Biometrika* **76** 425–433. MR1040637

Trotter, H. F. and Tukey, J. W. (1956). Conditional Monte Carlo for normal samples. In *Symposium on Monte Carlo Methods* (H. Meyer, ed.) 64–79. Wiley, New York. MR0079825

Ventura, L. (1997). Metodi asintotici, semiparametrici e robusti per l'inferenza in famiglie di gruppo. Ph.D. thesis, Dept. Statistics, Univ. Padova, Italy. (In Italian.)

Zamar, D., McNeney, B. and Graham, J. (2007). `elrm`: Software implementing exact-like inference for logistic regression models. *J. Statist. Software* **21**, issue number 3. Available at http://www.jstatsoft.org/v21/i03.